\documentclass[conference]{IEEEconf}

\input epsf
\usepackage{graphicx}
\usepackage{multirow}
\usepackage{titlesec}
\usepackage{balance} 
\usepackage{stfloats}
\usepackage{xcolor}
\usepackage{xurl}
\usepackage{booktabs}
\usepackage{subcaption}
\usepackage{tabularx}
\usepackage{hyperref}
\usepackage{xtab}
\usepackage{colortbl} 
\usepackage{xcolor}
\usepackage{array} 
\definecolor{lightblue}{RGB}{220,240,255} 
\definecolor{lightpink}{RGB}{255, 230, 230}
\definecolor{verylightgray}{RGB}{240, 240, 240}
\definecolor{verylightyellow}{RGB}{255, 255, 200}
\definecolor{b}{RGB}{173, 216, 230}
\definecolor{p}{RGB}{230, 230, 255}
\definecolor{u}{RGB}{242, 230, 242}
\usepackage{fdsymbol}
\usepackage{tikz}

\renewcommand\thesection{\arabic{section}} 

\renewcommand\thesubsectiondis{\thesection.\arabic{subsection}}

\renewcommand\thesubsubsectiondis{\thesubsectiondis.\arabic{subsubsection}}

\begin{document}

\title{\textbf{\Large Decoding Social Sentiment in DAO: A Comparative Analysis of Blockchain Governance Communities\\}}

\author{Yutong Quan$^{1}$, Xintong Wu$^{1}$, Wanlin Deng$^{1}$, and Luyao Zhang$^{1,*}$\\
	\normalsize $^{1}$Data Science Research Center, Duke Kunshan University, Suzhou, China\\

	\normalsize *The corresponding author: lz183@duke.edu
}


\maketitle
\begin{abstract}
Blockchain technology is leading a revolutionary transformation across diverse industries, with effective governance as a critical determinant for the success and sustainability of blockchain projects. Community forums, pivotal in engaging decentralized autonomous organizations (DAOs), wield a substantial impact on blockchain governance decisions. Concurrently, Natural Language Processing (NLP), particularly sentiment analysis, provides powerful insights from textual data. While prior research has explored the potential of NLP tools in social media sentiment analysis, a gap persists in understanding the sentiment landscape of blockchain governance communities. The evolving discourse and sentiment dynamics on the forums of top DAOs remain largely unknown. This paper delves deep into the evolving discourse and sentiment dynamics on the public forums of leading DeFi projects---\textit{Aave}, \textit{Uniswap}, \textit{Curve Dao}, \textit{Yearn.finance}, \textit{Merit Circle}, and \textit{Balancer}—placing a primary focus on discussions related to governance issues. Our study shows that participants in decentralized communities generally express positive sentiments during Discord discussions. Furthermore, there is a potential interaction between discussion intensity and sentiment dynamics, i.e., a higher discussion volume may contribute to a more stable sentiment from code analysis. The insights gained from this study are valuable for decision-makers in blockchain governance, underscoring the pivotal role of sentiment analysis in interpreting community emotions and its evolving impact on the landscape of blockchain governance. This research significantly contributes to the interdisciplinary exploration of the intersection of blockchain and society, specifically emphasizing the decentralized blockchain governance ecosystem.
We provide our data and code for replicability as open access on  GitHub\footnote{{\url{https://github.com/SciEcon/BlockchainSentiment2023}}}.
\end{abstract}
\IEEEoverridecommandlockouts
\vspace{1.5ex}
\begin{keywords}
\itshape blockchain governance, decentralized finance (DeFi), sentiment analysis, social media, Discord, natural language processing (NLP)
\end{keywords}

%
\IEEEpeerreviewmaketitle

\section{Introduction}
Blockchain technology has sparked a transformative wave in the financial landscape, marking the advent of a new era in decentralized finance (DeFi). This revolutionary technology has redefined the way capital flows and financial services are offered, fundamentally challenging traditional financial intermediaries. At its core, DeFi empowers individuals by eliminating the need for intermediaries~\cite{schueffel_defi:_2021}, thus enhancing financial inclusivity and security. Central to the DeFi movement are two key components: decentralized exchanges (DEXs) and DEBs. These innovative platforms represent a paradigm shift in how financial transactions and services are conducted.

\begin{itemize}
    \item \textbf{Decentralized Exchanges (DEXs).} DEXs stand at the forefront of DeFi, revolutionizing the trading and exchange of digital assets. In contrast to traditional centralized exchanges (CEXs), DEXs operate on blockchain networks, enabling users to trade directly from their wallets without relying on intermediaries~\cite{makridis_rise_2023,mohan_automated_2022,aspris_decentralized_2021}. This decentralized model offers several advantages, including (1) reduced counterparty risk by eliminating the need to trust a centralized exchange with private keys, (2) the potential for lower transaction fees, and (3) a more extensive range of trading pairs that provide access to riskier or less liquid cryptocurrencies~\cite{lin2019deconstructing}. Renowned DEXs such as \textit{Uniswap}, SushiSwap, and PancakeSwap have gained prominence for facilitating peer-to-peer trading while ensuring liquidity~\cite{yano_blockchain_2020,barbon_quality_2023,parlour_decentralized_2021,lo_dexs_2022,aspris_decentralized_2021,makridis_rise_2023}. Consequently, DEXs play a pivotal role in the DeFi ecosystem, ushering in a new era of accessibility and inclusivity in the realm of finance.
    \item \textbf{Decentralized banks (DEBs)} DEBs have emerged as a cornerstone of DeFi~\cite{guo2016blockchain,ao2022decentralized,zhang2023blockchain}. These platforms offer a range of financial services, including lending, borrowing, and earning interest on cryptocurrencies. Notably, \textit{Aave}, Compound, and MakerDAO are leading projects in this domain. DEBs leverage smart contracts to automate lending and borrowing processes, providing users with greater control over their assets. Users can lend their assets to earn interest or borrow assets by collateralizing their own, all without the need for traditional banks or credit checks. This innovative approach has the potential to disrupt traditional banking systems by offering borderless and inclusive financial services to a global audience.
\end{itemize}

In today's digital age, social media platforms serve as dynamic and influential hubs for shaping public opinion and fostering discourse. Within the blockchain space, these platforms play a pivotal role in connecting and engaging community members~\cite{tong2022people}. They serve as virtual town squares where participants share their insights, express their views, and collectively navigate the evolving blockchain landscape. Within these vibrant communities, social media serves crucial functions to influence governance decisions. In general, the intersection of blockchain technology and social media platforms creates a dynamic ecosystem where decentralized finance thrives. As blockchain governance decisions become increasingly important, understanding the emotions and opinions expressed within these communities becomes crucial for project leaders, developers, and stakeholders alike~\cite{fu2023ai,zhang2023mechanics}. This study aims to explore the emotional dynamics within blockchain governance communities by focusing on prominent DeFi projects—\textit{Uniswap}, \textit{Aave}, \textit{Curve Dao}, \textit{Yearn.finance}, \textit{Merit Circle}, and \textit{Balancer}. Selected based on their high market capitalization, active Discord forums, and possession of governance coins, these projects offer valuable insights into the sentiments and attitudes of their communities toward governance decisions. Furthermore, while Twitter was once a primary platform for social media sentiment analysis, the limitations posed by the closure of the Twitter API have made studying social media sentiment more challenging. However, given Discord's active role as a central communication platform for many blockchain governance communities and its offering of an open API for examining community discussions in this domain~\cite{rennie_toward_2022}, our research relies on discussion data from six DeFi communities on the Discord platform. This approach allows us to effectively capture the sentiments of blockchain governance participants regarding governance decisions. Specifically, we aim to answer the following research questions:

{\bfseries RQ1:} What common trends can be identified in the social sentiment across the leading DeFi protocols?

{\bfseries RQ2:} What are the differences in user engagement across leading DeFi protocols, and how is this engagement correlated with changes in sentiment?

{\bfseries RQ3:} What are the key topics discussed in the Discord communities of leading DeFi protocols, and how do these discussions influence sentiment scores?

By answering these research questions, our study aims to provide a comprehensive understanding of the complex discourses within blockchain governance communities, thereby making a valuable contribution to the growing body of knowledge in the field of blockchain governance. Our findings suggest that although these DeFi communities have different levels of discussion engagement on Discord, they all show a common trend of overall positive social sentiment. Section~\ref{data} describes the methods used for data collection, preprocessing, and the application of the VADER (Valence Aware Dictionary and Sentiment Reasoner) sentiment analysis tool in the context of blockchain sentiment analysis. In addition, Section \ref{results} presents our findings from the quantitative analysis of the social discourse in these different DeFi communities. Section~\ref{conclusion} concludes. In the Appendix, we present the background of relevant literature in Section~\ref{background}, offer a qualitative analysis to deepen the understanding of the quantitative results in Section~\ref{qua}, and discuss future research directions in Section~\ref{future}.

\section{Data and Method}
\label{data}

To conduct a comprehensive sentiment analysis of the top DeFi communities, we reviewed the top 200 coins ranked by market capitalization on CoinGecko\footnote{https://www.coingecko.com/en/categories/governance} under "Top Governance Coins by Market Cap" (data as of October 1, 2023). After careful examination, we found that out of the top 18 governance coins by market capitalization, 13 coins have Discord forums. To ensure the reliability of our analysis, we conducted a further screening of the forums' activity levels.

Initially, we checked whether each Discord community has dedicated channels specifically for discussing governance issues, with each channel having a discussion volume exceeding 500 messages. Notably, \textit{Uniswap}, \textit{Aave}, \textit{Curve Dao}, and \textit{Yearn.finance} have specific "governance" channels. 

However, \textit{Merit Circle} and \textit{Balancer} do not have dedicated channels for discussing governance issues. Despite this, we made a special case and reviewed the data within their respective "general" discussion channels for reference in our analysis. This rigorous selection process ensures that our analysis of these selected DeFi protocols is based on active and relevant discussions within the Discord community.

\subsection{Data Collection}

\begin{table*}[!ht]
\centering
\caption{DApp Discord Community Discussion Statistics (Data as of August 15, 2023)}
\label{tab:discussion-stats}
\begin{tabular}{@{}lccccl@{}}
\toprule
\textbf{DApp} & \textbf{Discord Channel} & \textbf{Start Date} & \textbf{End Date} & \textbf{Message Count} \\
\midrule
Uniswap & \#governance & September 17, 2020 & August 13, 2023 & 11,583 \\
Aave & \#governance & August 2, 2019 & July 27, 2023 & 3,844 \\
Curve DAO & \#curve-governance & August 18, 2020 & August 15, 2023 & 8,774 \\
yearn.finance & \#governance & July 20, 2020 & August 2, 2023 & 7,214 \\
\midrule
Merit Circle & \#general & February 14, 2022 & August 15, 2023 & 34,786 \\
Balancer & \#general & November 29, 2019 & August 15, 2023 & 51,707 \\
\bottomrule
\end{tabular}
\end{table*}

After careful examination, we gathered data from Discord forums. We employed the DiscordChatExporter\footnote{https://github.com/Tyrrrz/DiscordChatExporter} tool, which provides a streamlined method for accessing and exporting discussion data dating back to the inception of the channels.

As of August 15, 2023, our dataset includes discussions from the Discord communities of \textit{Uniswap}, \textit{Aave}, \textit{Curve Dao}, \textit{Yearn.finance}, \textit{Merit Circle}, and \textit{Balancer}. Table~\ref{tab:discussion-stats} provides specific details regarding each project's corresponding Discord channel, the start of discussions, the end of discussions, and the total volume of discussions. 

The dataset encompasses various essential variables that facilitate comprehensive analysis:

\begin{itemize}
\item{\verb|AuthorID|}: This identifier uniquely distinguishes the authors of the discussions, allowing for tracking and attribution.
\item{\verb|Author|}: The name or username of the discussion participants.
\item{\verb|Date|}: The timestamp indicates when each discussion occurred, providing a temporal dimension to the dataset.
\item{\verb|Content|}: The textual content of the discussions, including messages, comments, and replies.
\item{\verb|Attachment|}: Information regarding any attached files, images, or media shared within the discussions.
\item{\verb|Reactions|}: A record of reactions, such as emojis, associated with each discussion, offering insights into community engagement and sentiment.
\end{itemize}

By collecting data from these Discord channels from top DeFi communities, we have built a valuable dataset that provides a comprehensive understanding of the discussions, interactions, and sentiments expressed by community members. This dataset serves as the basis for our sentiment analysis and further examination of the sentiments and attitudes prevalent in these blockchain governance communities.

\subsection{Data Preprocessing}

Before performing sentiment analysis on the data, the textual content must be preprocessed to eliminate noise, irrelevant information, and inconsistencies from the textual data to ensure the accuracy and significance of the subsequent analysis. The process of data preprocessing includes lowercasing, URL removal, special character removal, tokenization, stop word removal, and punctuation removal in order to clean and structure the raw text data for optimal analysis. In our study, the main focus is on the "Content" column, which contains the textual content of discussions from the Discord channel. After the preprocessing steps described above, the cleaned and structured text data is stored in a new column labeled "Preprocessed". This column contains the preprocessed text ready for sentiment analysis.

\subsection{Sentiment Analysis}

In this study, sentiment analysis plays a pivotal role in uncovering the sentiment dynamics within the discussions held in the Discord "governance" channels of the top DeFi communities. Sentiment analysis is an NLP technique utilized to discern, extract, and quantify subjective information from textual data, such as social media posts and online discussions~\cite{pang2008opinion,liu2012sentiment}.

To perform sentiment analysis, we leveraged the VADER (Valence Aware Dictionary and Sentiment Reasoner) sentiment analysis tool. VADER is a rule-based model explicitly designed for sentiment analysis of social media text~\cite{hutto_vader_2014}, making it particularly suitable for analyzing the informal and dynamic nature of discussions in these Discord channels.

VADER employs a lexicon that assigns sentiment scores to lexical features within the text~\cite{hutto_vader_2014}. These scores capture the sentiment expressed in each piece of text and are then aggregated to generate a compound sentiment score~\cite{hutto_vader_2014}. The compound score falls within a range from -1 to 1, where -1 signifies extremely negative sentiment, 1 represents highly positive sentiment, and scores approaching zero indicate a neutral sentiment stance~\cite{hutto_vader_2014}.

In our analysis, we applied the VADER sentiment analysis tool to each discussion within the "governance" channels of  \textit{Uniswap}, \textit{Aave}, \textit{Curve Dao},
\textit{Yearn.finance}, and the "general" channels of \textit{Merit Circle}, and \textit{Balancer} on Discord. This process generated a compound sentiment score for each discussion, reflecting the overall emotional tone expressed in these messages.

Given the large size of the dataset, instead of calculating the sentiment score for each message one by one, we performed separate date-based and user-based sentiment analyses. In the date-based sentiment analysis, we calculated the daily average sentiment score, which effectively represents the prevailing sentiment in the governance discussion on a given day. In user-based sentiment analysis, we calculated the average sentiment score for each user. For both dimensions of sentiment analysis, we also provide statistics on the results of the analysis accordingly, including minimum, maximum, variance, mean, etc., to comprehensively explore and compare changes in the sentiment of the discourse within the Top DeFi community.

\section{Results}
\label{results}

\subsection{Common Trends in Social Sentiment}
\subsubsection{Date-based Sentiment Analysis}
\begin{figure*}[ht]
    \centering
    
    \begin{subfigure}{.5\textwidth}
        \includegraphics[width=\linewidth]{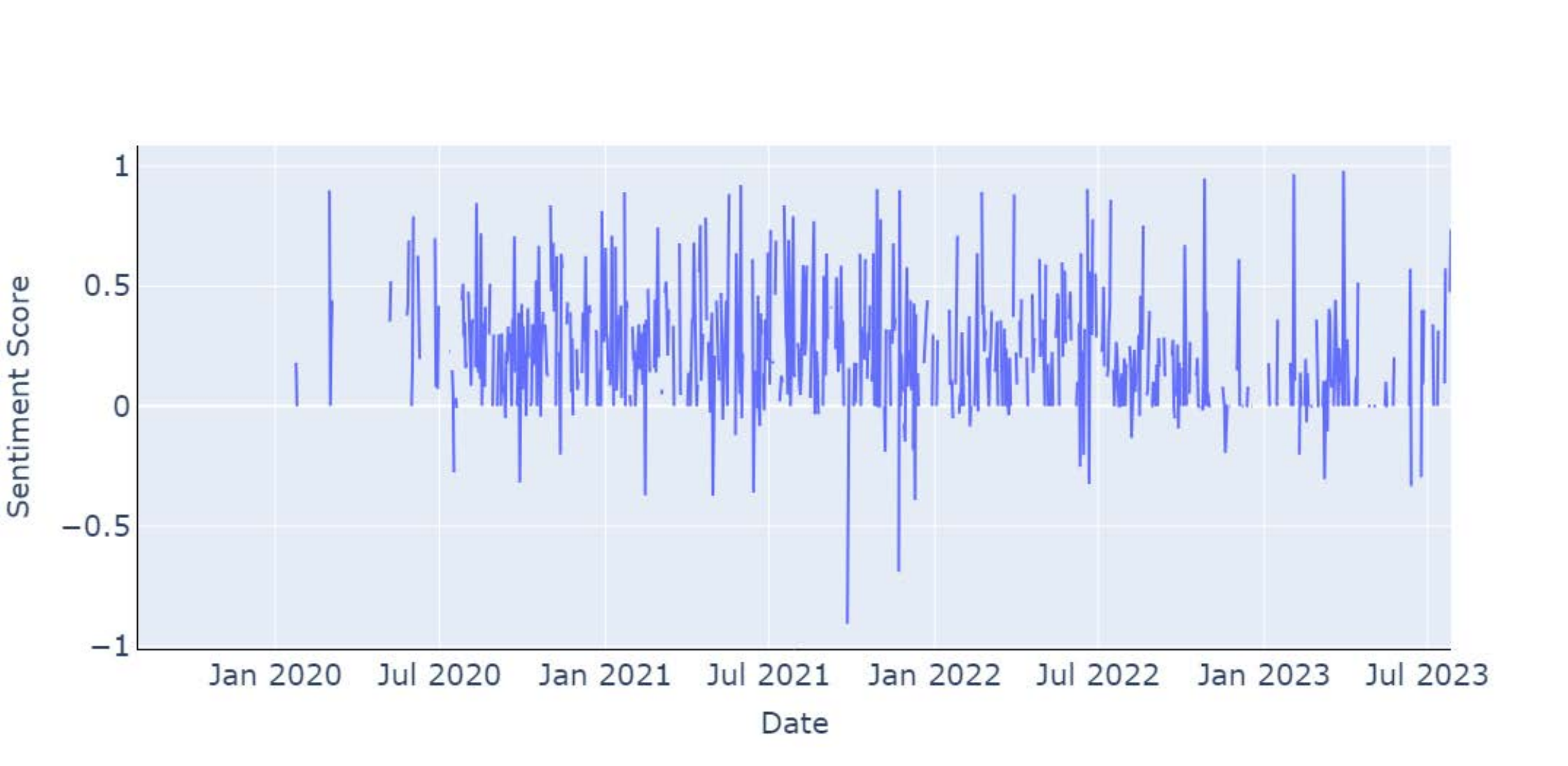}
        \caption{Aave}
        \label{fig:aavescore}
    \end{subfigure}%
    \begin{subfigure}{.5\textwidth}
        \includegraphics[width=\linewidth]{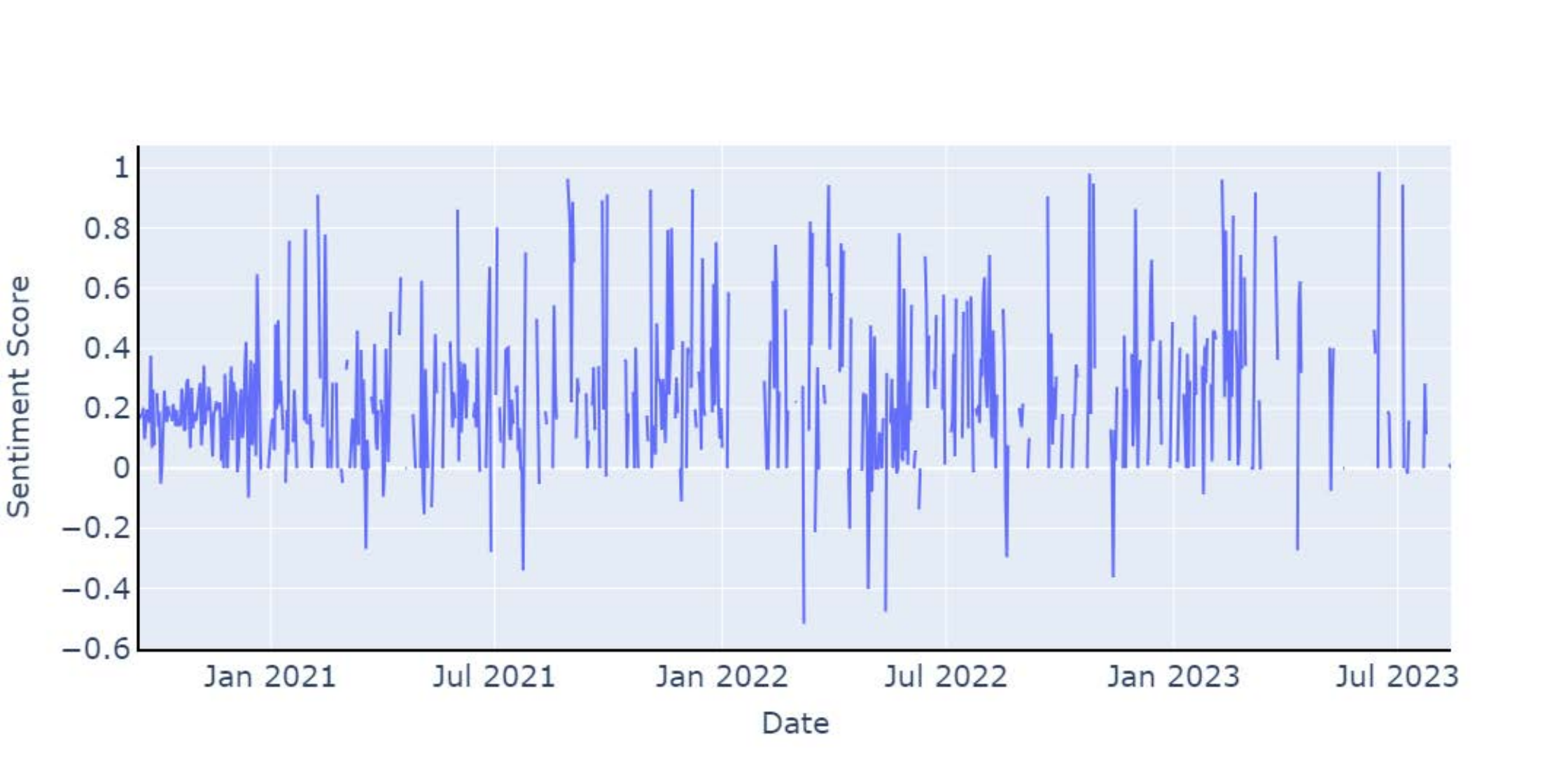}
        \caption{Uniswap}
        \label{fig:uniscore}
    \end{subfigure}

    \medskip

    \begin{subfigure}{.5\textwidth}
        \includegraphics[width=\linewidth]{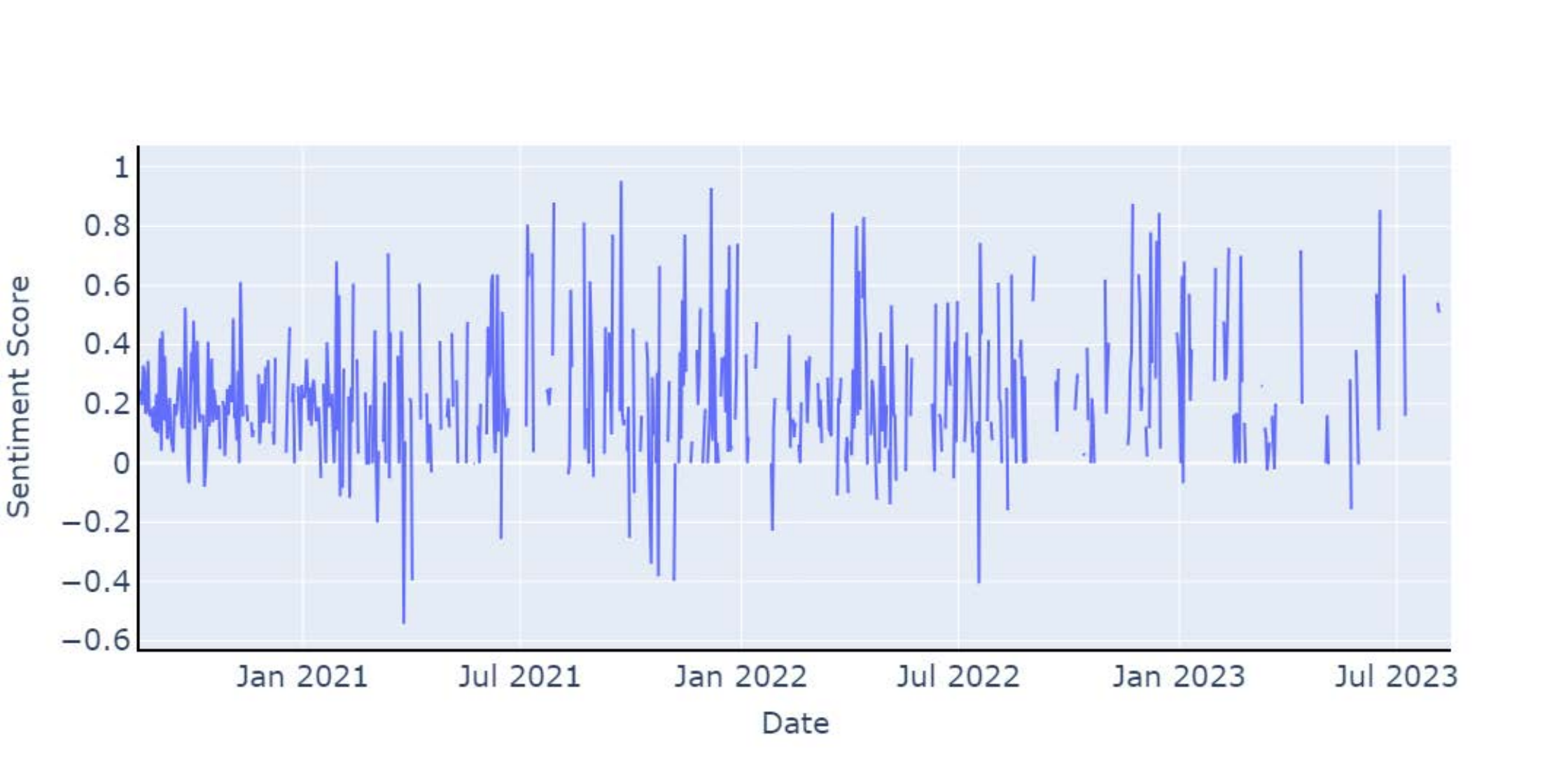}
        \caption{Curve Dao}
        \label{fig:curvescore}
    \end{subfigure}%
    \begin{subfigure}{.5\textwidth}
        \includegraphics[width=\linewidth]{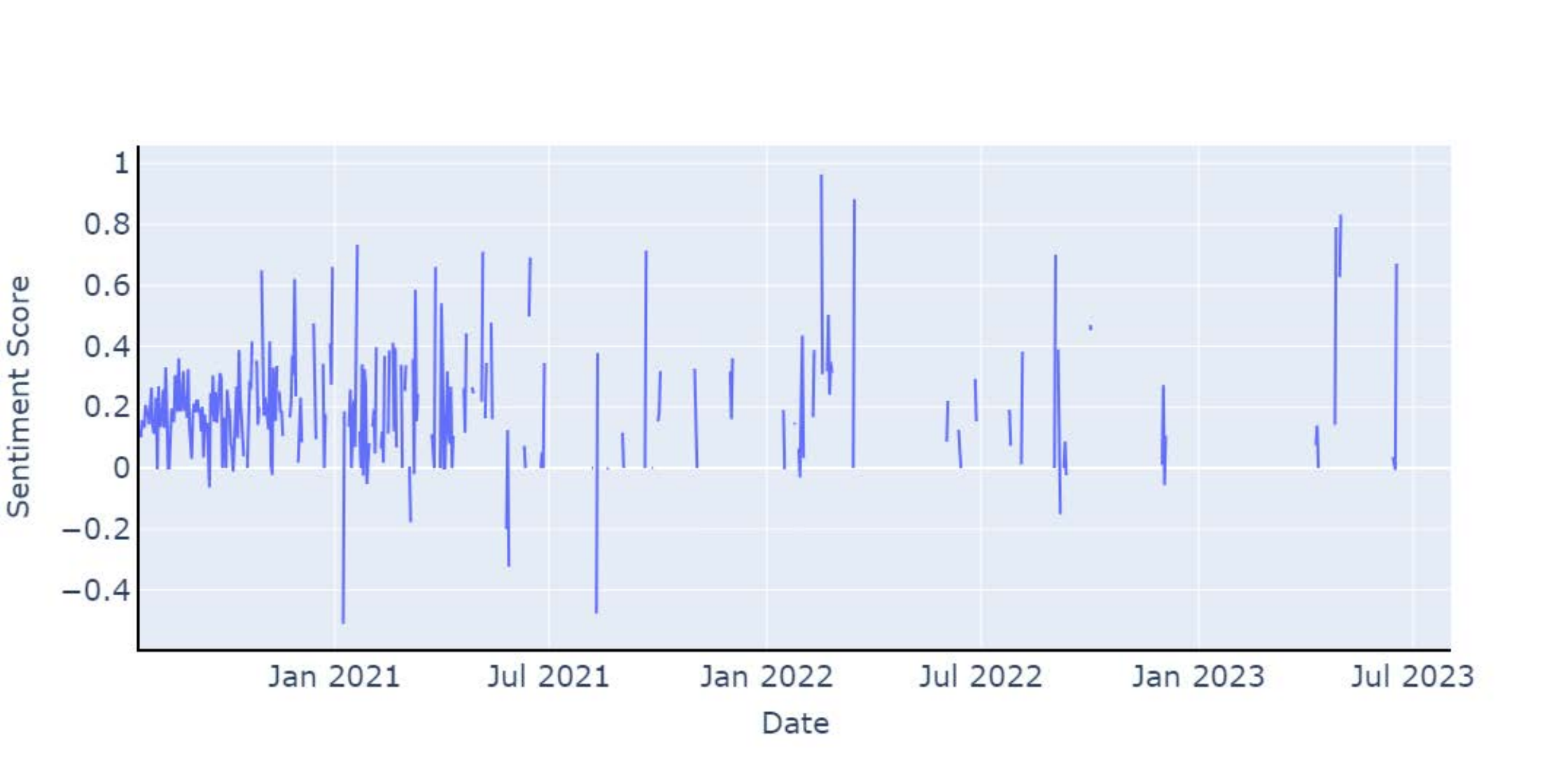}
        \caption{Yearn.finance}
        \label{fig:yearnscore}
    \end{subfigure}%
    
    \medskip
    
    \begin{subfigure}{.5\textwidth}
        \includegraphics[width=\linewidth]{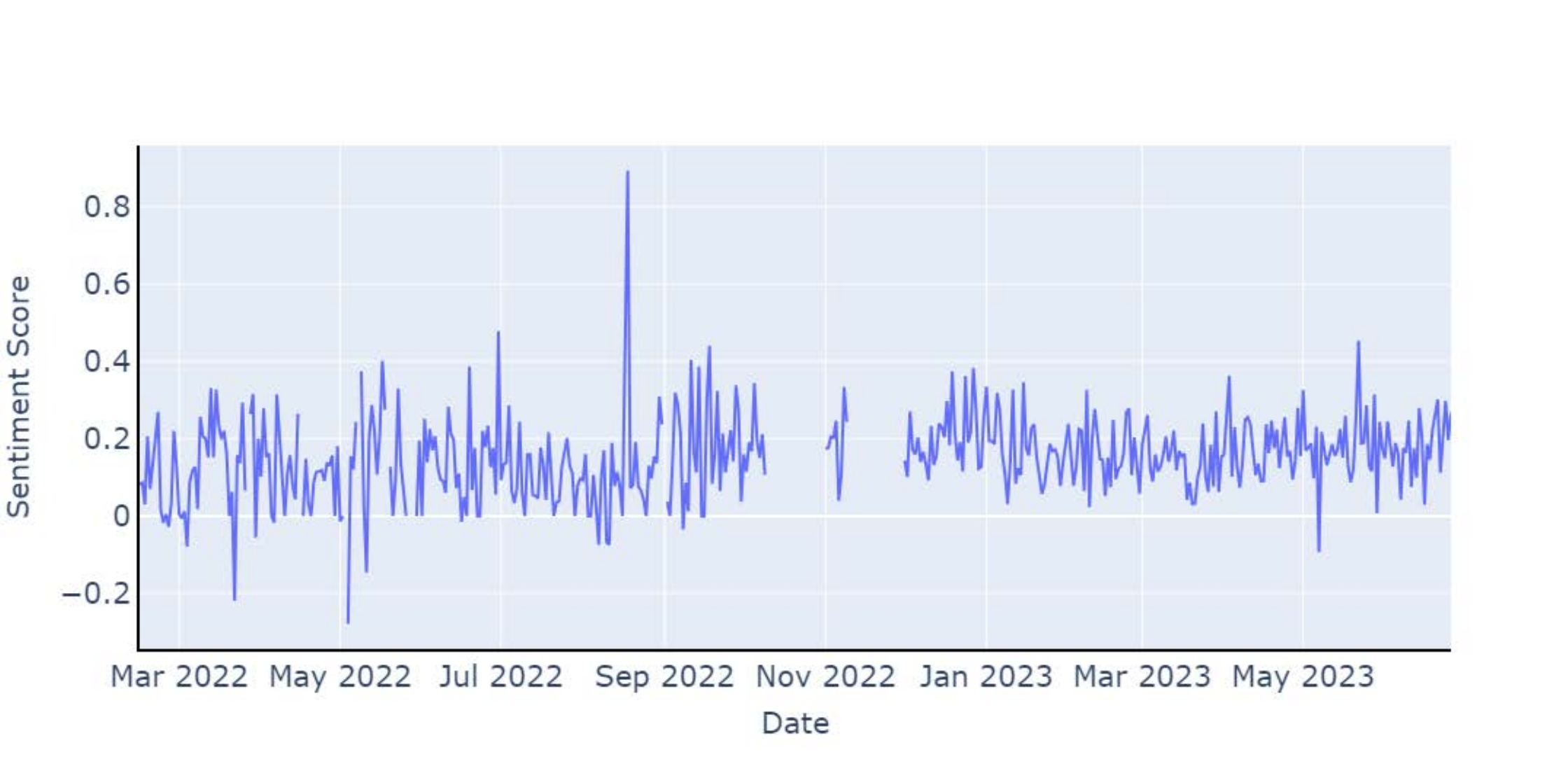}
        \caption{Merit Circle}
        \label{fig:meritscore}
    \end{subfigure}%
    \begin{subfigure}{.5\textwidth}
        \includegraphics[width=\linewidth]{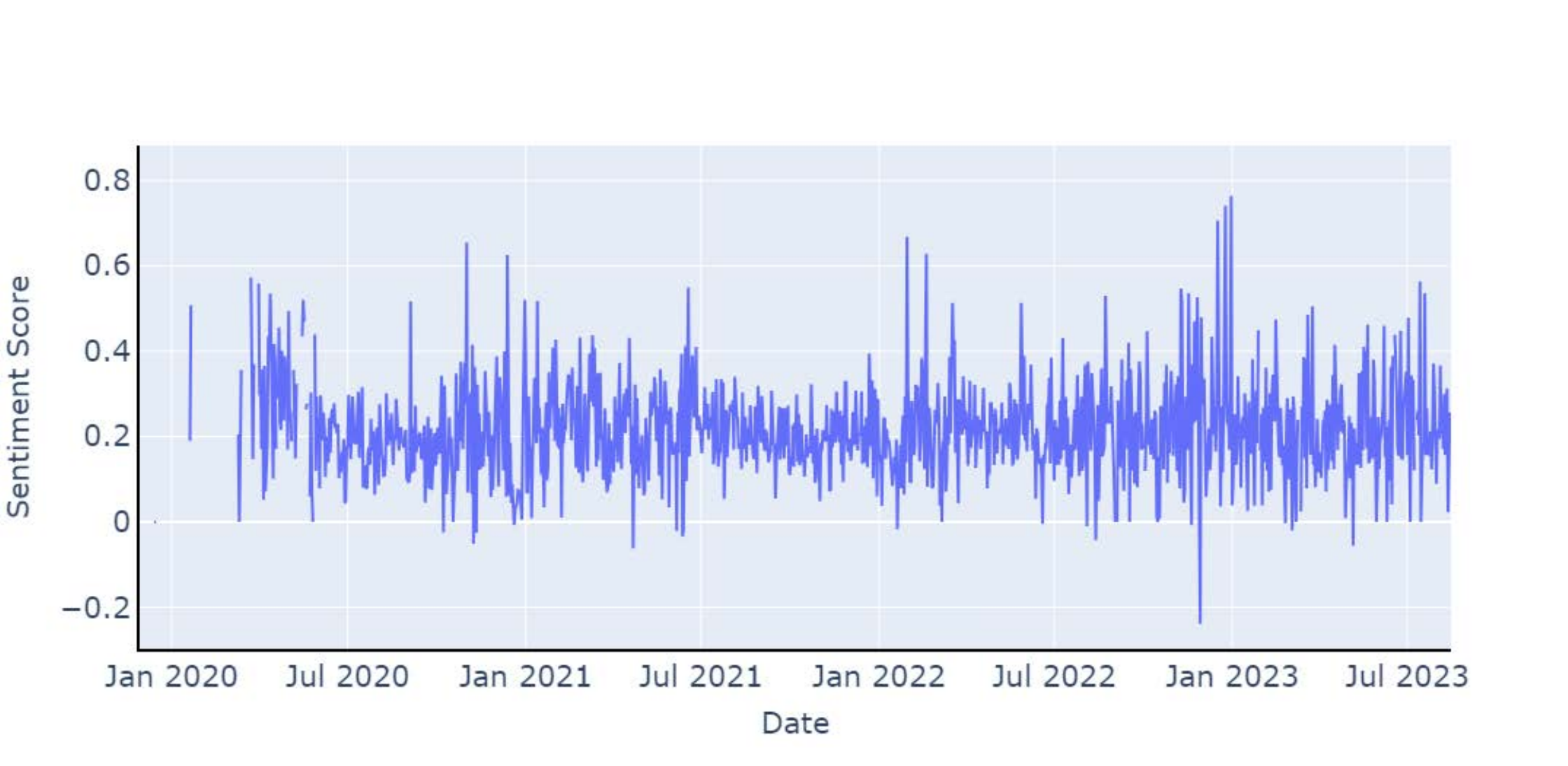}
        \caption{Balancer}
        \label{fig:bascore}
    \end{subfigure}
     
    \caption{Daily average sentiment scores in Discord discussions of the top DeFi protocols.}
    \label{fig:fullscore}
\end{figure*}

Discussions in the governance community within the DeFi ecosystem show common trends in social sentiment. Our sentiment analysis includes six prominent DeFi protocols—\textit{Aave}, \textit{Uniswap}, \textit{Curve Dao}, \textit{Yearn.finance}, \textit{Merit Circle}, and \textit{Balancer}—reveal a generally positive sentiment in their community discussions.

In Figure~\ref{fig:fullscore}, we show a composite illustration of the average daily sentiment scores of Discord discussions in the top DeFi communities. Each subgraph corresponds to a specific protocol that captures dynamic sentiment changes over time. Across all subgraphs, the sentiment score line mostly fluctuates above the zero line indicating that positive sentiment is prevalent in all communities.

To validate the graph-based naked-eye observations, we also statistically analyzed the date-based sentiment analysis results.
\begin{table*}[!ht]
\centering
\caption{Comparison of Social Sentiment Statistics Among DApp Discord Community}
\label{tab:sentiment-comparison}
\begin{tabularx}{\textwidth}{Xcccccc}
\toprule
\textbf{DApp} & \textbf{Days} & \textbf{Mean} & \textbf{Std. Dev.} & \textbf{Min} & \textbf{Max} & \textbf{Positive (\%)} \\
\midrule
Aave & 610 & 0.114 & 0.207 & -0.906 & 0.9805 & 91.64 \\
Uniswap & 576 & 0.1539 & 0.2323 & -0.5171 & 0.9913 & 92.86 \\
Curve DAO & 561 & 0.1261 & 0.2009 & -0.5171 & 0.9874 & 92.66 \\
Yearn.finance & 301 & 0.0626 & 0.1503 & -0.5106 & 0.9758 & 92.69 \\
\midrule
Merit Circle & 473 & 0.1417 & 0.1116 & -0.2787 & 0.8934 & 96.40 \\
Balancer & 1223 & 0.1978 & 0.1247 & -0.2384 & 0.8225 & 98.62 \\
\bottomrule
\end{tabularx}
\end{table*}

Table~\ref{tab:sentiment-comparison} shows that all communities have more than 90\% of positive sentiment with a variance of less than 0.2, which represents generally stable and positive attitudes held by participants in Discord communities across all protocols reviewed.

\subsubsection{User-based Sentiment Analysis}
\begin{table*}[!htbp]
\centering
\caption{DApp Discord Community Discussion User Average Sentiment Scores Distribution (Data as of August 15, 2023)}
\label{tab:user-score}
\begin{tabularx}{\textwidth}{Xcccc}
\toprule
\textbf{Dapp} & \multicolumn{4}{c}{\textbf{Sentiment Score Range (User Proportion)}} \\
\cmidrule(lrrr){2-5}
& \textbf{[-1.0, -0.5)} & \textbf{[-0.5, 0.0)} & \textbf{(0.0, 0.5)} & \textbf{[0.5, 1.0)} \\
\midrule
Aave & 2.77\% & 10.44\% & 63.13\% & 23.65\% \\
Uniswap & 2.04\% & 10.98\% & 68.97\% & 18.01\% \\
Curve Dao & 0.00\% & 2.31\% & 87.66\% & 10.03\% \\
yearn.finance & 0.15\% & 5.66\% & 81.97\% & 12.22\% \\
\midrule
Merit Circle & 0.06\% & 18.87\% & 76.37\% & 4.70\% \\
Balancer & 0.36\% & 7.89\% & 83.10\% & 8.65\% \\
\bottomrule
\end{tabularx}
\end{table*}

Regarding the user-based sentiment analysis, Table \ref{tab:user-score} gives the average sentiment scores for each user in the Discord discussions of the six DeFi protocols (excluding users with consistently zero sentiment scores). We provide users' distribution across the range of sentiment scores in each community. By comparing the six communities, we find that all communities show a more pronounced contrast between the percentage of optimistic and pessimistic users. The majority of users participating in the blockchain governance discussion remain generally optimistic. The average sentiment scores of the users are mainly concentrated in the range of 0 to 0.5, indicating a neutral-positive attitude.

\subsection{Differences in Discussion Engagement}

In addition to the overall sentiment commonality, we also performed simple statistics and visualization of user engagement and discussion volume in the Discord community for the six protocols to see whether user engagement is correlated with sentiment change.

\subsubsection{User Data Analysis}
\begin{table*}[!htbp]
\centering
\caption{DApp Discord Community Discussion User Statistics (Data as of August 15, 2023)}
\label{tab:user-message}
\begin{tabularx}{\textwidth}{Xccccccc}
\toprule
\textbf{DApp} & \textbf{User Count} & \textbf{25\%} & \textbf{Median} & \textbf{75\%} & \textbf{Min} & \textbf{Max} & \textbf{Mean Messages per User}\\
\midrule
Aave & 830 & 1 & 2 & 3 & 1 & 589 & 4.63 \\

Uniswap & 1001 & 1 & 2 & 5 & 1 & 1226 & 11.57 \\

Curve Dao & 409 & 2 & 4 & 12 & 1 & 889 & 21.45 \\

yearn.finance & 722 & 1 & 3 & 11 & 1 & 647 & 18.84 \\
\midrule
Merit Circle & 2287 & 1 & 3 & 8 & 1 & 1827 & 15.21 \\

Balancer & 5270 & 1 & 2 & 6 & 1 & 2798 & 9.81 \\
\bottomrule
\end{tabularx}
\end{table*}

Table~\ref{tab:user-message} lists user-message statistics for the six DeFi protocols in the Discord forums. These statistics include the number of participating users, the distribution of user messages, and the average number of messages per user. We use these data to measure the level of participation in governance discussions within each community.

Our findings show that of the four main communities with "governance" channels, \textit{Uniswap}, \textit{Aave}, and \textit{Yearn.finance} have the highest number of users, suggesting that users are widely involved in governance matters. In contrast, although \textit{Curve Dao} has a smaller number of users, it has the highest average discussion participation, closely followed by \textit{Yearn.finance} and \textit{Uniswap}. As for the two reviewed community discussions in the "general" channel, the \textit{Balancer} shows higher user participation, while the \textit{Merit Circle} has a higher average discussion participation rate.

\subsubsection{Discussion Volume Analysis}

\begin{figure*}[ht]
    \centering
    
    \begin{subfigure}{.5\textwidth}
        \includegraphics[width=\linewidth]{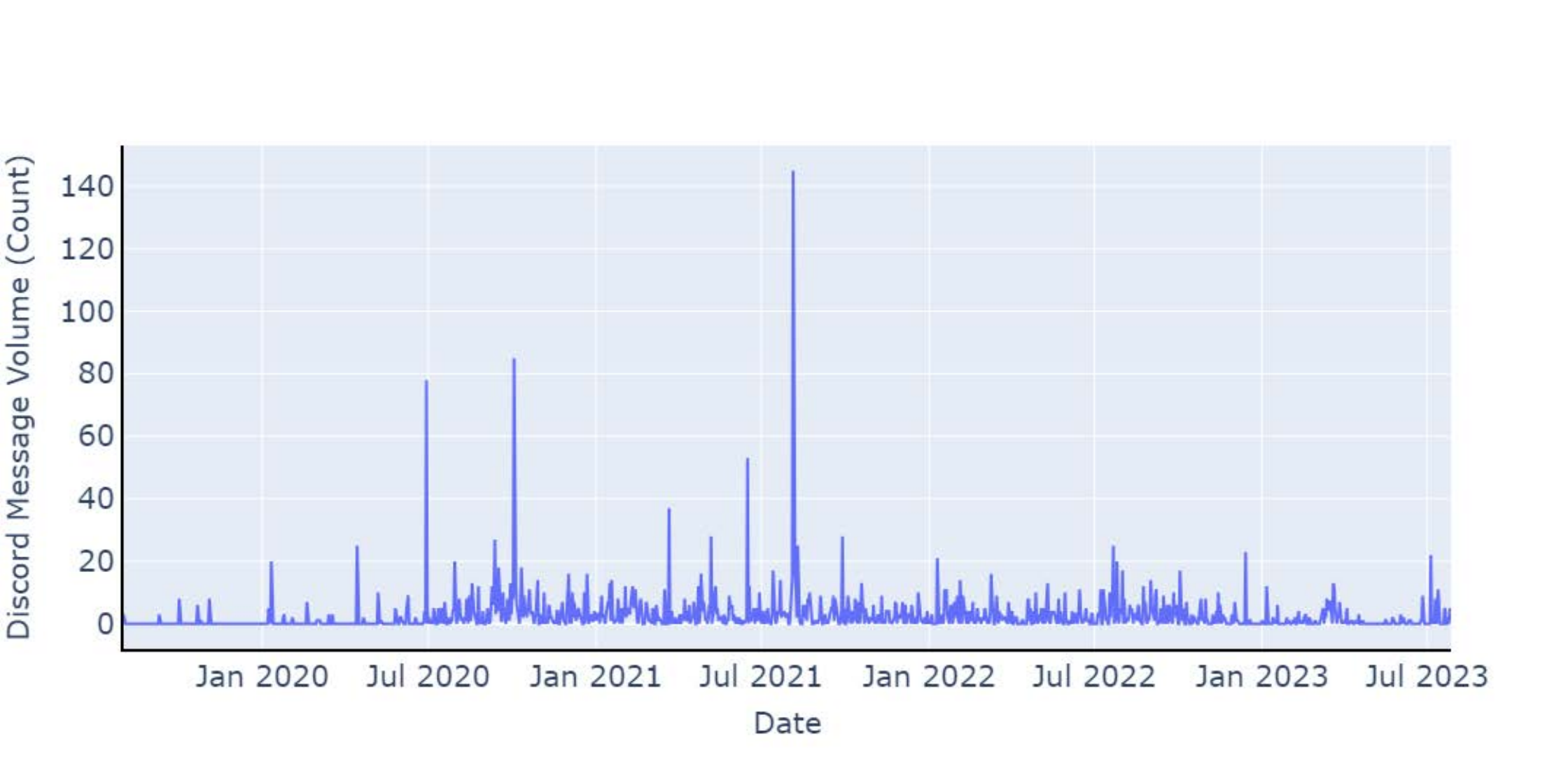}
        \caption{Aave}
        \label{fig:aavevolume}
    \end{subfigure}%
    \begin{subfigure}{.5\textwidth}
        \includegraphics[width=\linewidth]{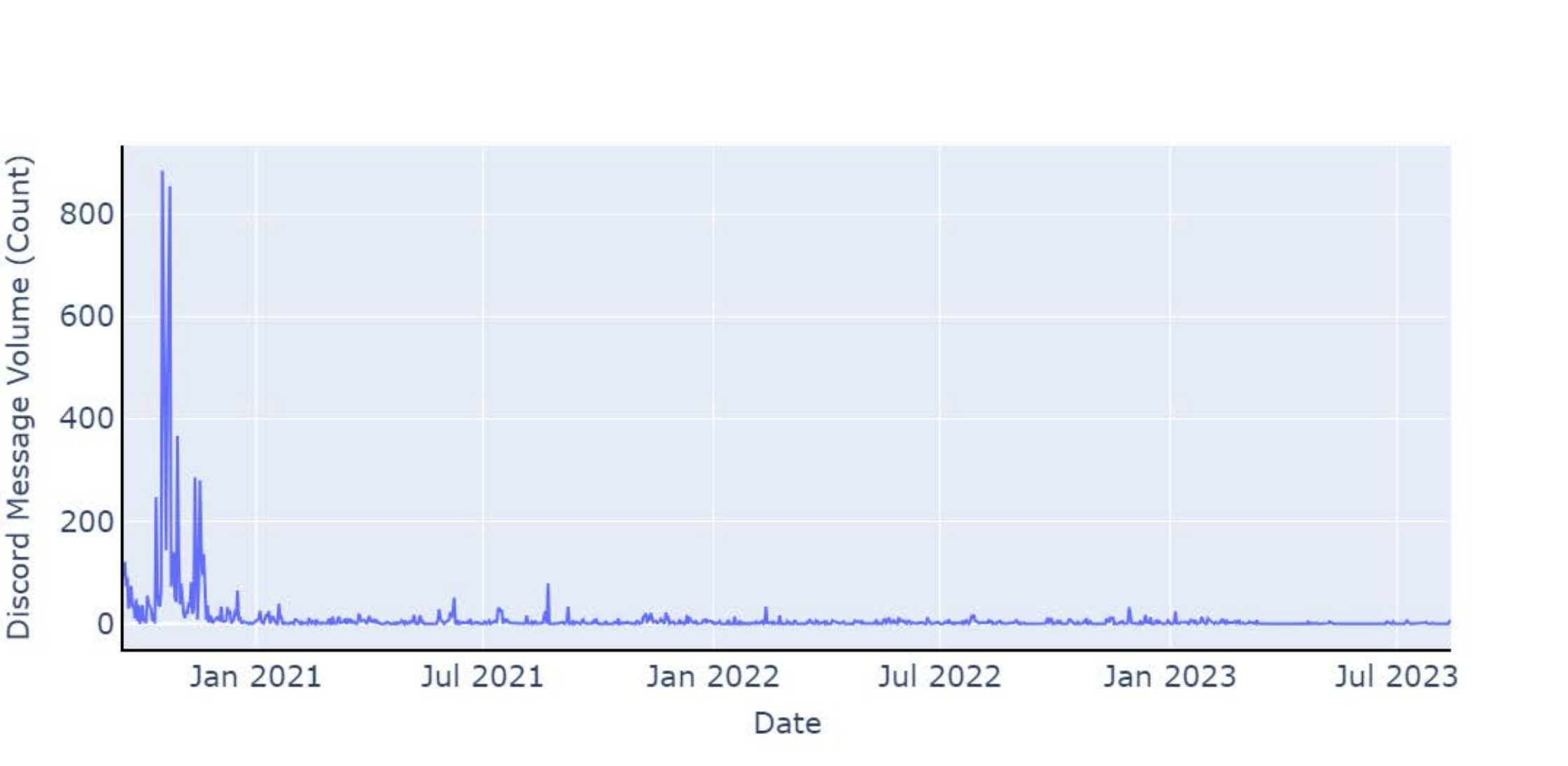}
        \caption{Uniswap}
        \label{fig:univolume}
    \end{subfigure}

    \medskip

    \begin{subfigure}{.5\textwidth}
        \includegraphics[width=\linewidth]{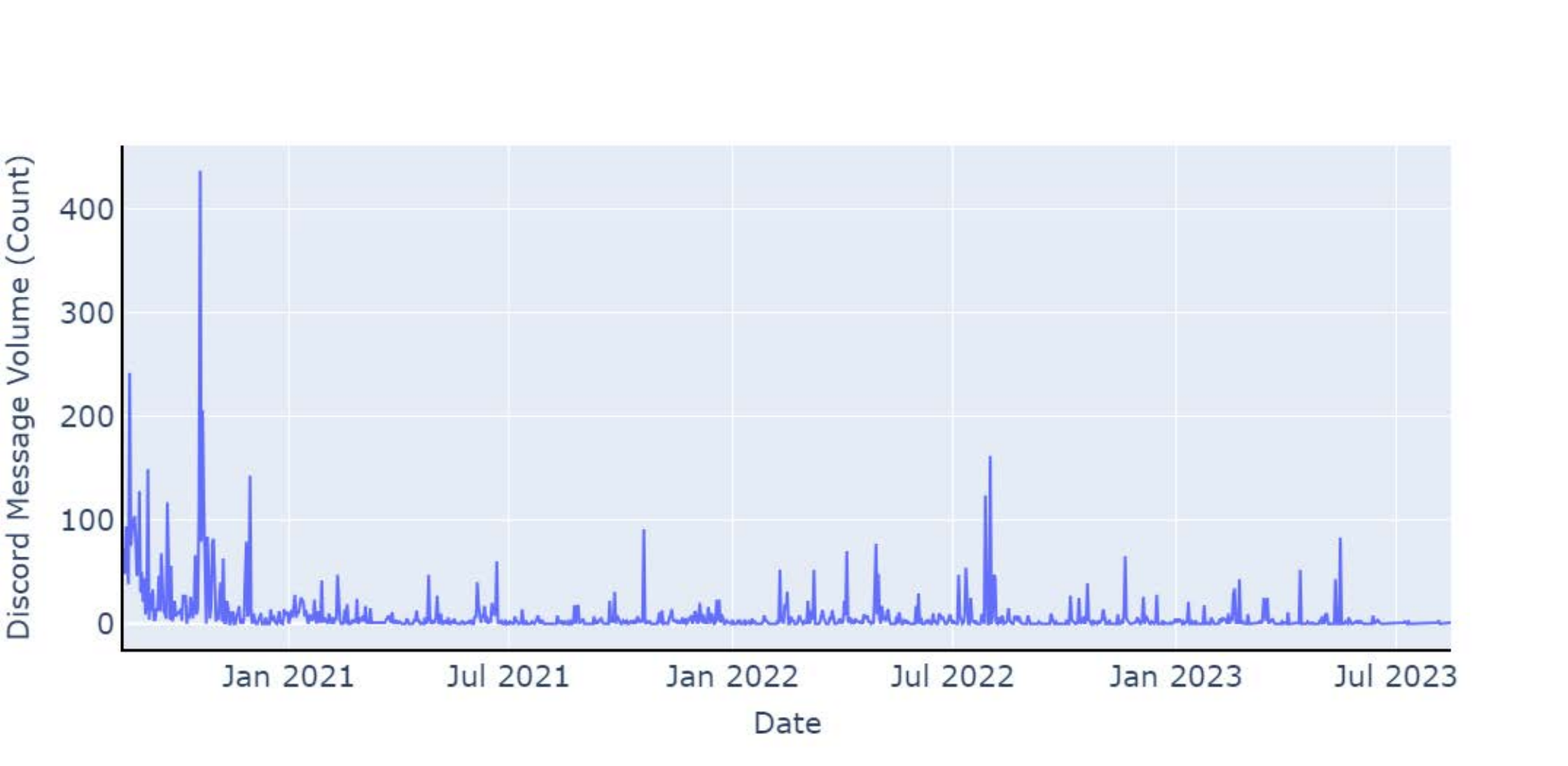}
        \caption{Curve Dao}
        \label{fig:curvevolume}
    \end{subfigure}%
    \begin{subfigure}{.5\textwidth}
        \includegraphics[width=\linewidth]{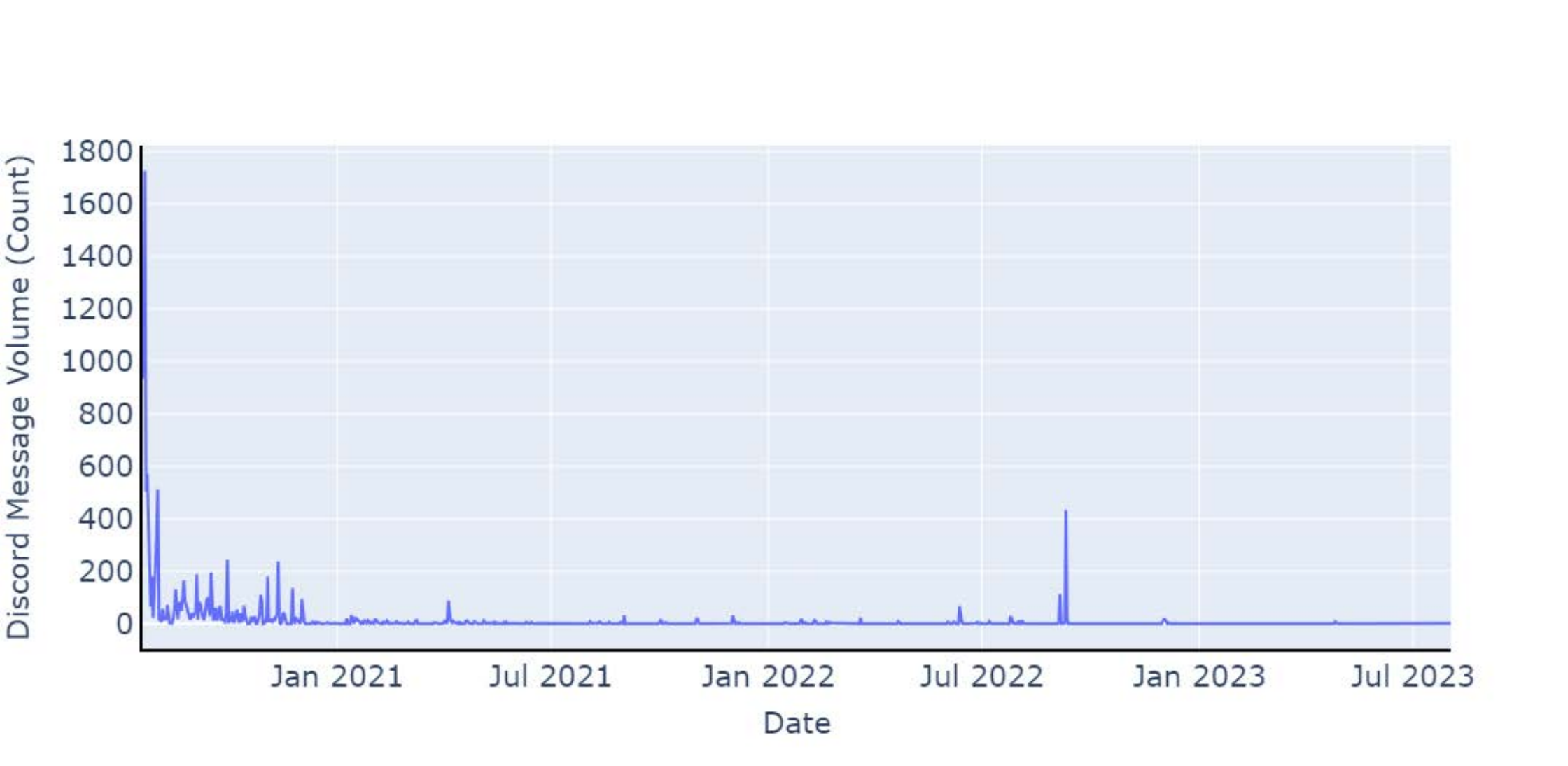}
        \caption{Yearn.finance}
        \label{fig:yearnvolume}
    \end{subfigure}%
    
    \medskip
    
    \begin{subfigure}{.5\textwidth}
        \includegraphics[width=\linewidth]{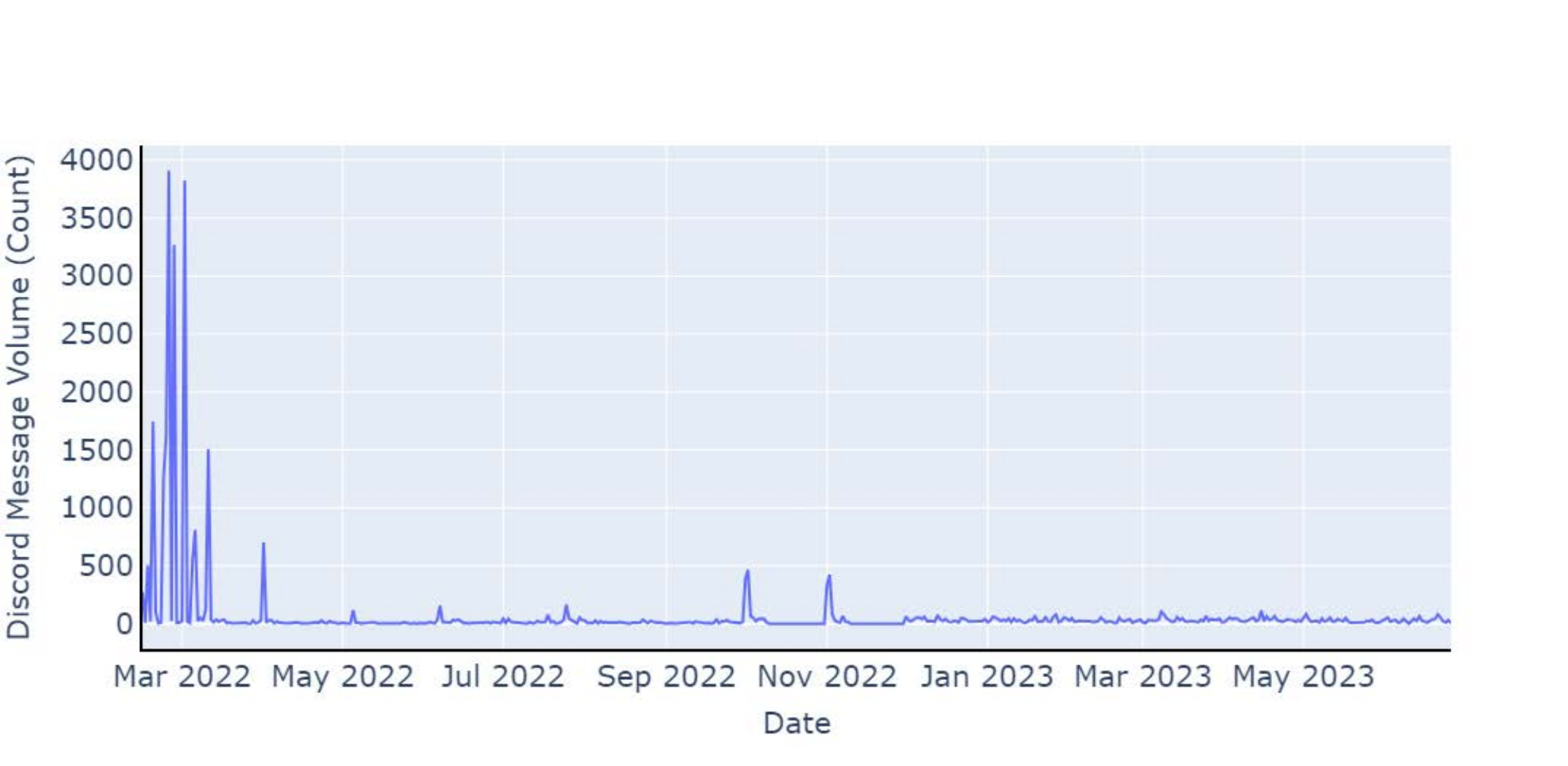}
        \caption{Merit Circle}
        \label{fig:meritvolume}
    \end{subfigure}%
    \begin{subfigure}{.5\textwidth}
        \includegraphics[width=\linewidth]{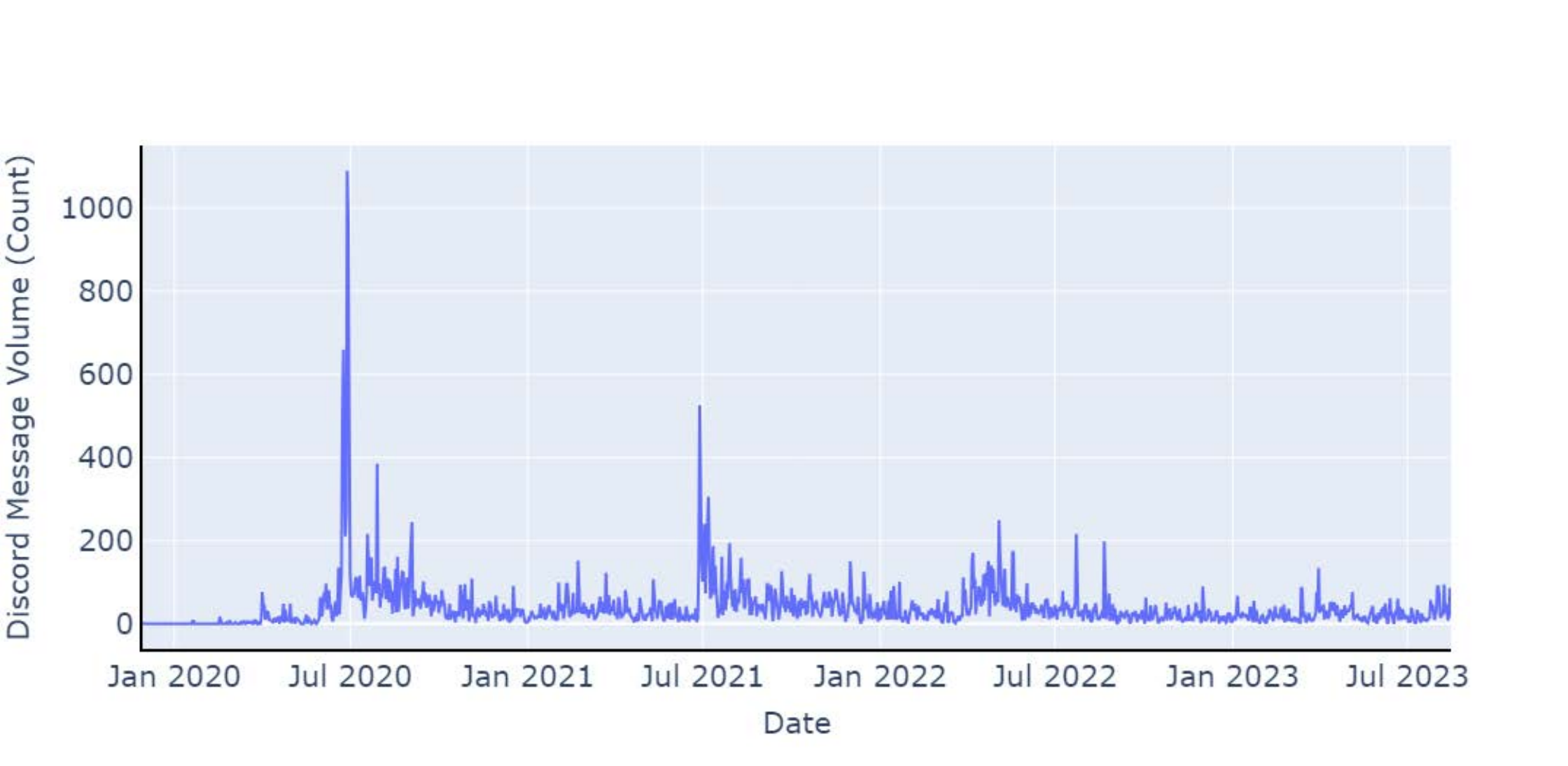}
        \caption{Balancer}
        \label{fig:bavolume}
    \end{subfigure}

    \caption{Daily Discord discussion volume of the top DeFi protocols.}
    \label{fig:fullvolume}
\end{figure*}

\begin{figure*}[ht]
    \centering
    
    \begin{subfigure}{.5\textwidth}
        \includegraphics[width=\linewidth]{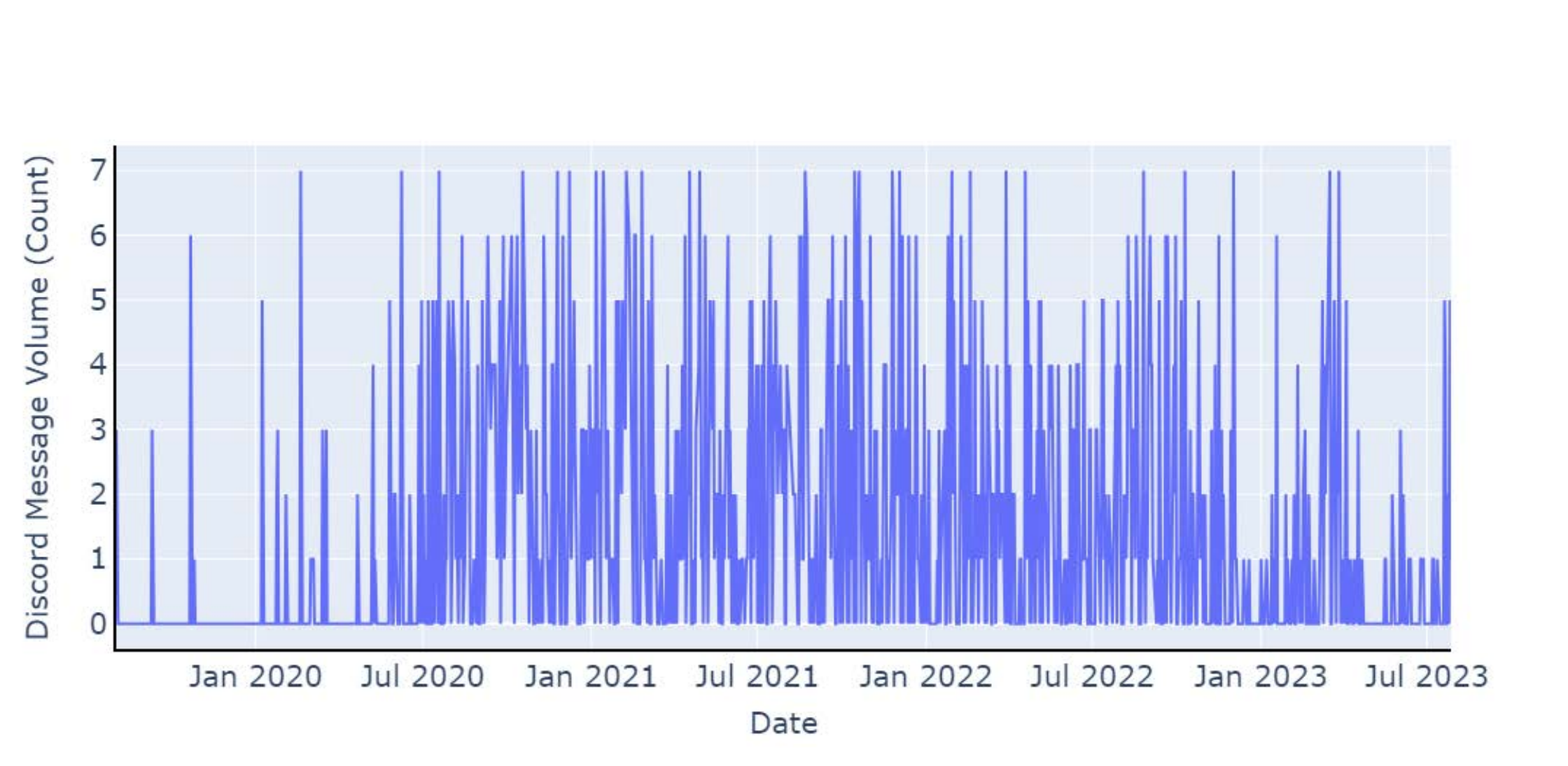}
        \caption{Aave}
        \label{fig:aavevol}
    \end{subfigure}%
    \begin{subfigure}{.5\textwidth}
        \includegraphics[width=\linewidth]{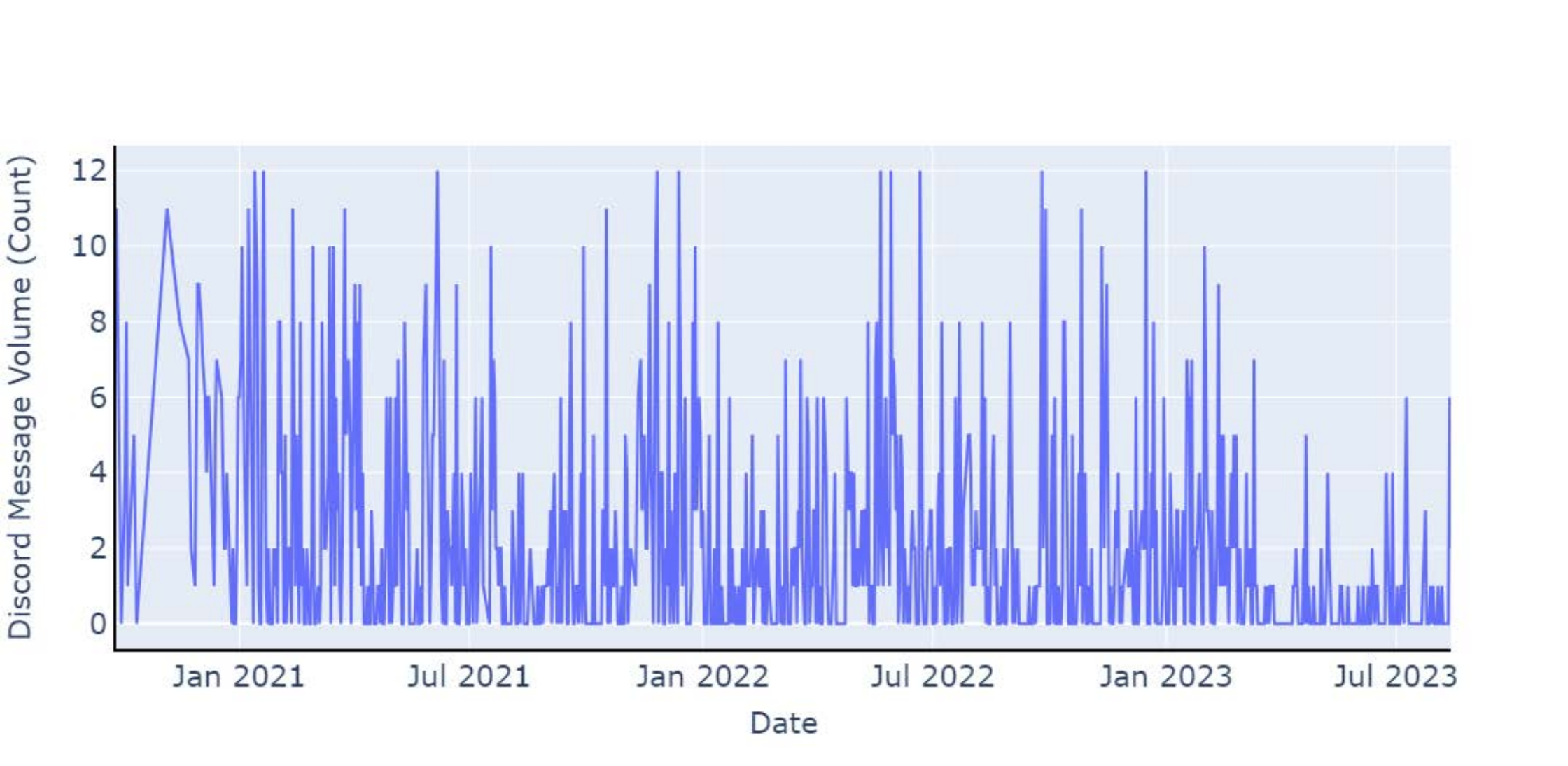}
        \caption{Uniswap}
        \label{fig:univol}
    \end{subfigure}

    \medskip

    \begin{subfigure}{.5\textwidth}
        \includegraphics[width=\linewidth]{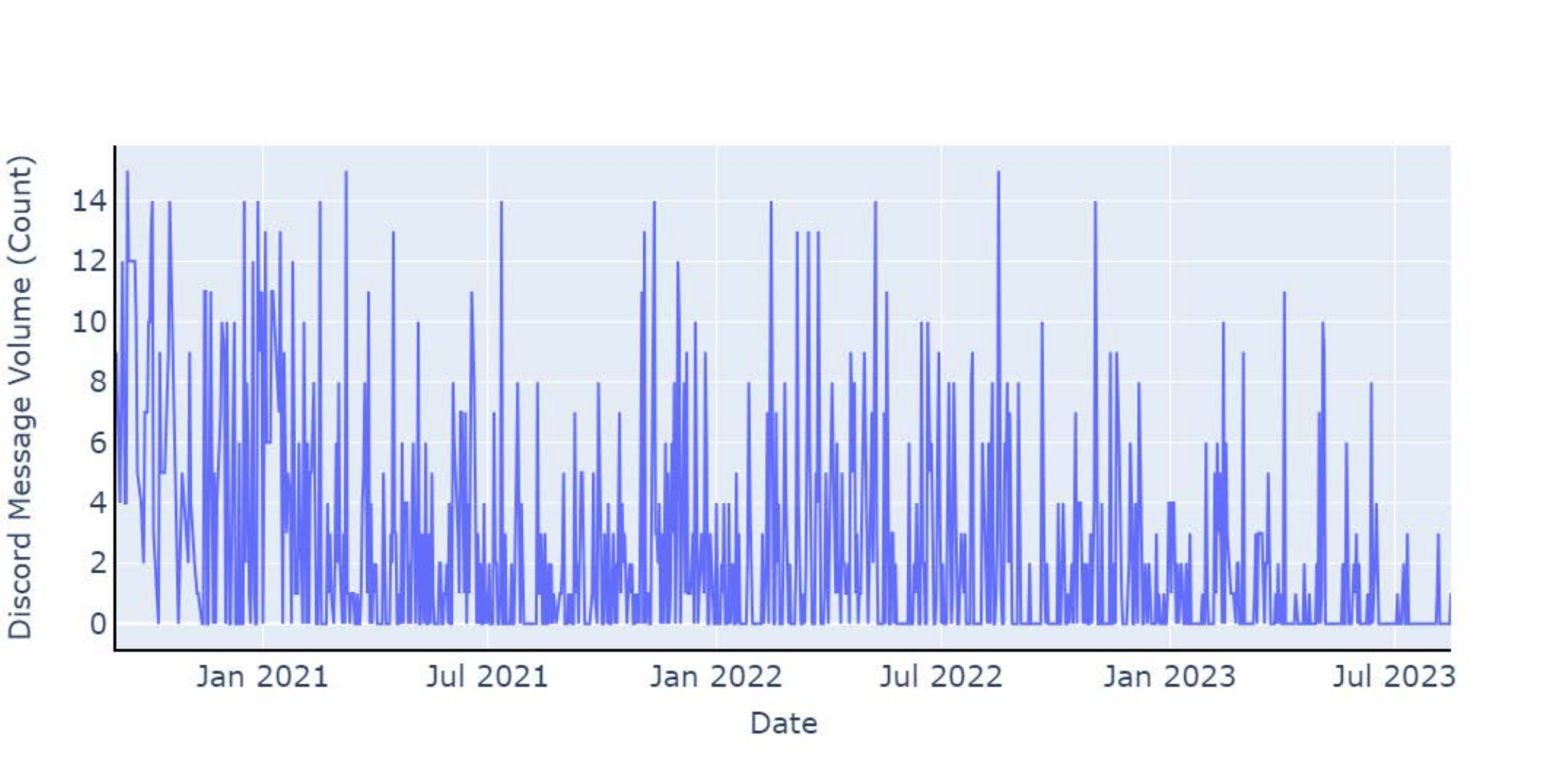}
        \caption{Curve Dao}
        \label{fig:curvevol}
    \end{subfigure}%
    \begin{subfigure}{.5\textwidth}
        \includegraphics[width=\linewidth]{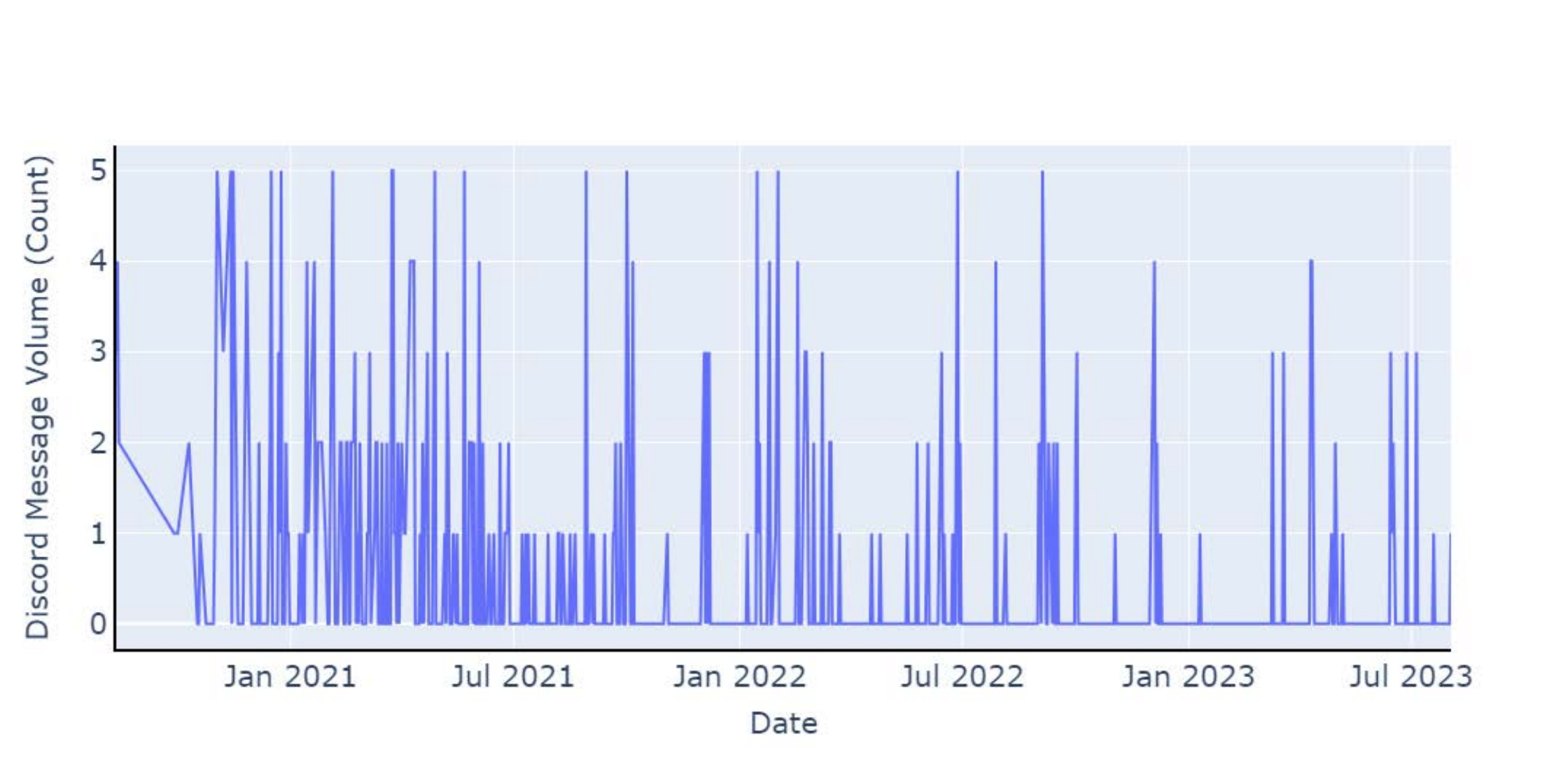}
        \caption{Yearn.finance}
        \label{fig:yearnvol}
    \end{subfigure}%
    
    \medskip
    
    \begin{subfigure}{.5\textwidth}
        \includegraphics[width=\linewidth]{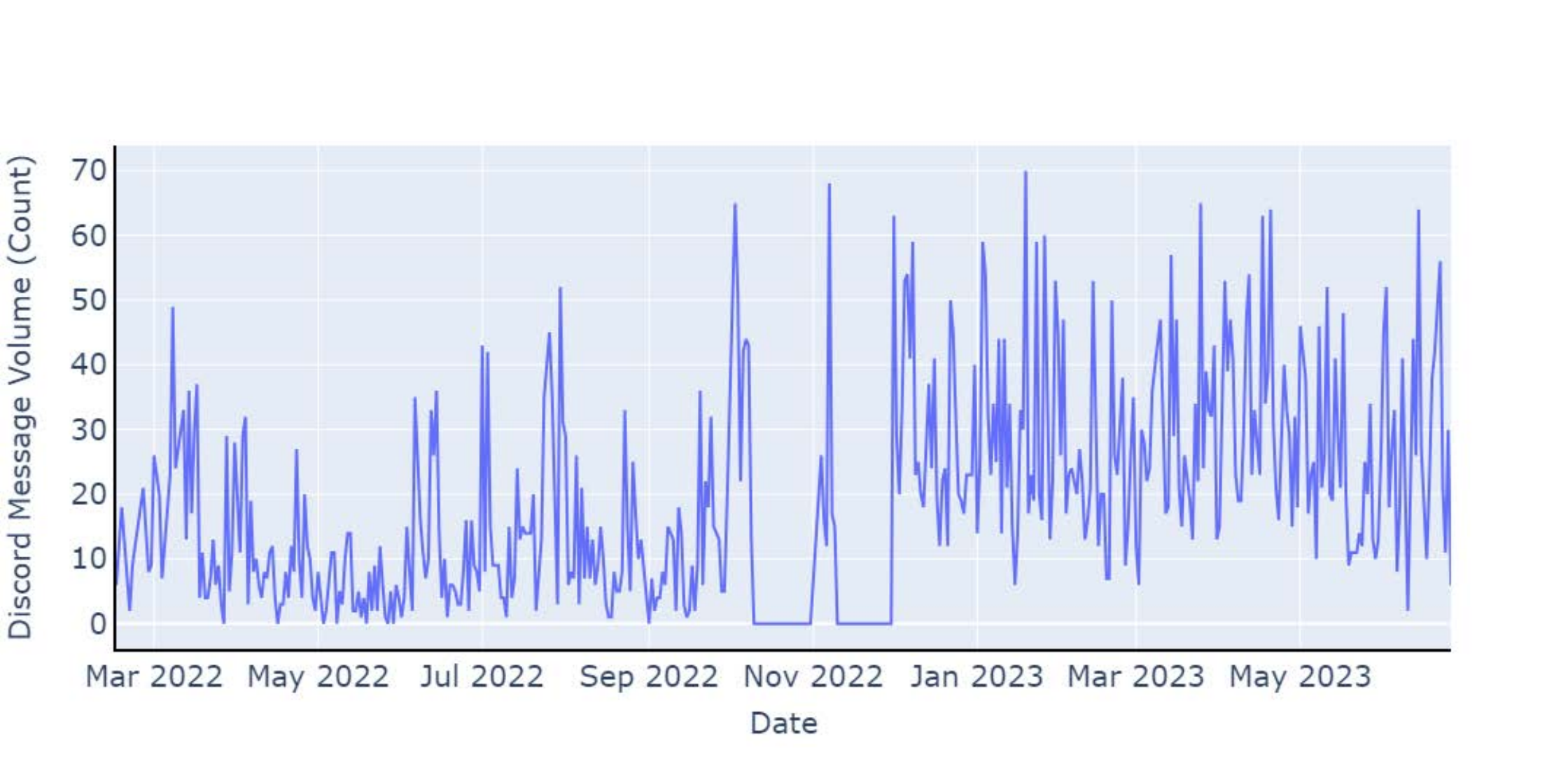}
        \caption{Merit Circle}
        \label{fig:meritvol}
    \end{subfigure}%
    \begin{subfigure}{.5\textwidth}
        \includegraphics[width=\linewidth]{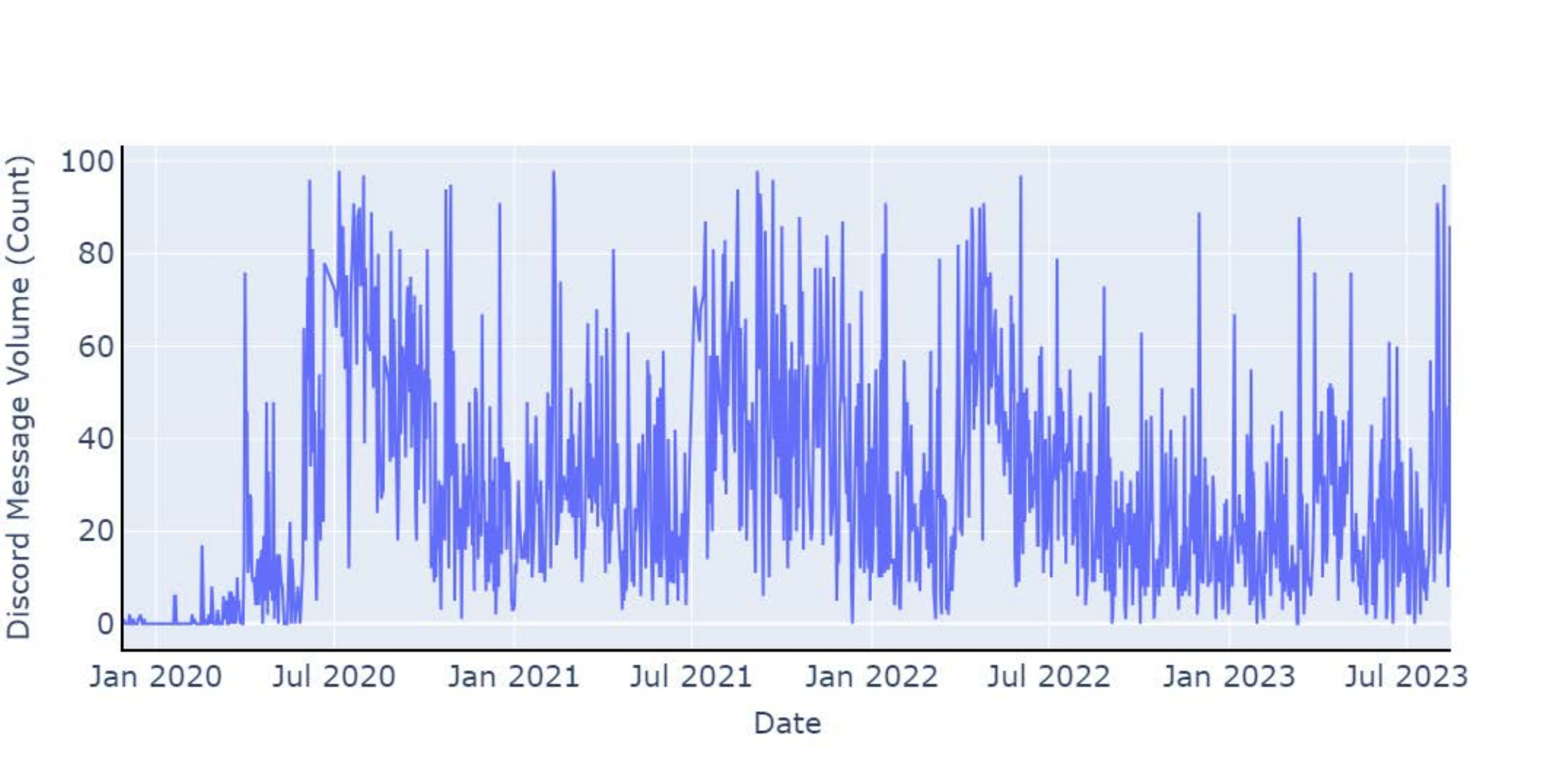}
        \caption{Balancer}
        \label{fig:bavol}
    \end{subfigure}

    \caption{Daily Discord discussion volume (outliers cleaned) of the top DeFi protocols.}
    \label{fig:cleanedvolume}
\end{figure*}

To explore potential correlations between discussion intensity and sentiment dynamics, we examined the daily discussion volume of dedicated Discord channels of each DeFi protocol. Figure \ref{fig:fullvolume} gives a time series graph illustrating the discussion volume of various DeFi protocols. It includes data from the start of their respective Discord discussions through August 15, 2023. However, the clarity of the data fluctuations of the non-outliers in the graph is affected by the unusually high discussion volume on certain days. Therefore, we created a cleaned statistical line graph (see Figure \ref{fig:cleanedvolume}).

In Figure~\ref{fig:cleanedvolume}, we have cleaned the daily average discussion data for the six DeFi protocols using the Interquartile Range (IQR) method for outlier detection. The IQR is a measure of statistical dispersion, calculated as the difference between the 75th percentile (Q3) and the 25th percentile (Q1) of the data. Lower and upper bounds are established for the dataset to define a range considered "normal." These bounds are computed using the IQR:

\begin{itemize}
\item{\verb|Lower bound|}: Q1 - 1.5 * IQR.
\item{\verb|Upper bound|}: Q3 + 1.5 * IQR.
\end{itemize}

Subsequently, data points below the lower limit or above the upper limit were considered outliers and removed, leaving only the "normal" values within the upper and lower limits to observe specific fluctuations in the amount of discussion.

When we delved into the interplay between discussion intensity and mood dynamics, we found that protocols such as \textit{Uniswap}, \textit{Curve Dao}, and \textit{Yearn.finance}, marked by higher discussion intensity in their initial phases, consistently showed more stably positive sentiment scores in the corresponding periods. This interesting pattern suggests a potential correlation, indicating that higher discussion intensity corresponds to a more stable emotional tone in the community. To further confirm this correlation, we can examine the \textit{Balancer} "general" channel, which has a wider range of discussions. The apparent increase in discussions after July 2020 coincides with a significant stabilization of sentiment. This additional observation by the \textit{Balancer} community is consistent with common trends in \textit{Uniswap}, \textit{Curve Dao}, and \textit{Yearn.finance}, suggesting the possibility that the increased intensity of the discussion is associated with a more stable sentiment. 

\subsection{Discussion Themes}

In addition to the sentiment review, we also explored what was specifically being discussed in these governance channels and how this discussion content influenced the examined sentiment results. Overall, for the four governance channels, two themes dominated the discussion: (1) a significant amount of the discussion was about problems users encountered in the processes of governance, proposals, and voting, such as \textit{"How many aave we need to hold for voting?"}, \textit{"Do I need to hold Aave tokens to submit AIP?"}, and \textit{"How can I create a proposal to have governance?"}; (2) There are a fair number of third-party advertisements posted in the hope of gaining users' attention or reaching out to DeFi protocol officials.

In \textit{Aave}'s forum, we noticed an official message, \textit{"gm fam, Many of the questions asked here lately are being redirected to the governance forums. Therefore it makes sense that this channel will be read-only starting today. If you have any governance questions we encourage you to search and ask on the governance forums. Thanks! https://governance.aave.com/."} This message may explain why there isn't more discussion about governance mechanisms in Discord's governance forums, which are mostly centered around governance processes. More review of the discussion on governance mechanisms may need to be done in the official governance forums of \textit{Aave} in future studies.

In \textit{Curve Dao}'s forums, we were pleasantly surprised to see more discussions about actual governance strategies. For example, on September 4, 2020, a user posted \textit{"Has there been a proposal or discussions to change gauge weight voting? I'm thinking whether changing CRV emission I'm thinking whether changing CRV emission weights to be in proportion to fee generation would be more beneficial to Curve."} offered his thoughts on the new proposal and provided a description. Other users followed up with replies, such as \textit{"this is actually a good one for proposal (I'm not sure about tech feasibility yet) but rewards based on volume utilization makes sense?"} The sentiment score given by the machine for the day's discussion was 0.42, which is honestly hard to notice in general observation, being sandwiched in the middle of positive and neutral. However, we think such a classification can be recognized as a neutral-positive attitude, as users expressed their approval of the proposal idea while offering many ideas and thoughts for improvement.

The \textit{Yearn.finance} forum had also seen discussions of possible future proposals, such as between June 14 and June 16, 2023, when a user posted a discussion about security and risk \textit{"gm! who can I chat with regarding security and risk mgmt within Yearn"} and proposed a new technology, \textit{"Hypernative uses a combination of machine learning, heuristics, and real-time monitoring to predict when an attack is about to happen. Our machine learning models are trained on historical data to learn patterns of both legitimate and fraudulent transactions. Heuristics are rules-based systems that identify potential attacks that may not be detected by machine learning models and real-time monitoring collects data from a variety of sources to identify potential attacks."} But unfortunately, that discussion did not continue.

Therefore, our manual review showed that the dominant discussion in the four Discord governance channels centered around governance process issues, followed by advertising messages posted by third parties. Even if some discussion about governance strategies and future proposals did exist, this information was usually scarce, and the discussion was not actively pursued. This indicates a direction for future research, which is to conduct a further review of the official governance channels of the individual DeFi protocols, where there may be more discussion and opinions on governance strategies.

\section{Conclusion}
\label{conclusion}

\subsection{Key Findings}

Our in-depth analysis of social sentiment within blockchain governance communities focuses on \textit{Aave}, \textit{Uniswap}, \textit{Curve Dao}, \textit{Yearn.finance}, \textit{Merit Circle}, and \textit{Balancer}, providing valuable insights into the sentiment dynamics of DeFi protocols. Sentiment analysis of discussions in these different communities, review of participation, and examination of discussion content enrich our understanding of how participants engage in decentralized governance.

\subsubsection{Common Positive Sentiment Trends} 
A common theme has emerged in the Discord communities of all the DeFi protocols reviewed - the prevalence of positive sentiment. In the \textit{Aave}, \textit{Uniswap}, \textit{Curve Dao}, \textit{Yearn.finance}, \textit{Merit Circle} and \textit{Balancer} communities, participants maintained a generally positive sentiment in Discord discussions.

\subsubsection{User Engagement and Sentiment Stability}
Different protocols differ in their user engagement and trends in discussion volume over time. However, the results of the comparative analyses suggest a potential correlation whereby more intensive user engagement and discussion would allow for more stable sentiment results from code analysis.



\section*{Acknowledgment}

Prof. Luyao Zhang is supported by the National Science Foundation China (NSFC) on the project entitled “Trust Mechanism Design on Blockchain: An Interdisciplinary Approach of Game Theory, Reinforcement Learning, and Human-AI Interactions (Grant No. 12201266).” This research is also supported by the Center for Study of Contemporary China (CSCC) Undergraduate Research Grant.  Yutong Quan and Xintong Wu are supported by the pioneering Summer Research Scholar Program and Wang-Cai Biochemistry Lab Donation Fund at Duke Kunshan University in Summer 2023 for Prof. Luyao Zhang's related project titled "From 'Code is Law' to 'Code and Law': A Comparative Study on Blockchain Governance."

\balance
\bibliographystyle{ieeetr}
\bibliography{references}

\newpage
\appendix

\section*{Background}
\label{background}

Previous studies related to natural language processing (NLP) in sentiment analysis, blockchain governance, and DAOs have laid the groundwork for the exploration of the research questions in this study. NLP aims to achieve tasks such as text disambiguation, language translation, question-answering, and deriving meaningful inferences from text. Sentiment analysis, a crucial aspect of NLP, involves automatically identifying emotions and emotional polarity in text. It is widely employed for analyzing various textual materials, particularly social media text, to reflect user emotions and shifts in public sentiment. This study utilizes sentiment analysis through NLP as a research method to perform sentiment analysis within blockchain communities. Specifically, we employ a tool called VADER (Valence Aware Dictionary and Sentiment Reasoner), which is widely used in social media sentiment analysis.

Subsequently, we delve into the realm of blockchain governance. Previous research has underscored the significance of blockchain governance in unleashing the potential of blockchain transformation. It has provided a comprehensive framework for understanding the definition, classification, and patterns of blockchain, with an emphasis on the importance of social governance within blockchain spaces. Building upon this, we utilize blockchain forums as an application scenario to investigate the emotional dynamics within blockchain governance communities, offering valuable insights for decision-makers in social governance.

Finally, we focus on decentralized autonomous organizations (DAOs), representing a pinnacle of decentralized hierarchical management structures within blockchain governance. Given that DAOs are one of the most symbolic implementations of blockchain governance, this paper's application scenario concentrates on several top DeFi protocols' DAOs to study the sentiment dynamics within blockchain governance. Previous research has also discussed the inherent limitations of decentralized mechanisms in DAOs, highlighting the need for more comprehensive social analysis to promote participation and fairness in DAO governance.

\subsection{Natural Language Processing (NLP) in Sentiment Analysis}
Natural Language Processing (NLP) is a theoretically motivated range of computational techniques for analyzing and representing naturally occurring texts at one or more levels of linguistic analysis to achieve human-like language processing for a range of tasks or applications~\cite{liddy2001natural}. As an application at the intersection of computer science, artificial intelligence, and linguistics, NLP aims to achieve text paraphrasing, language translation, text comprehension for question answering, and drawing meaningful inferences from text. 

Sentiment analysis (SA), as an important task in natural language processing, aims to automatically identify the sentiment and sentiment polarity in a text, which usually includes positive, negative, and neutral sentiments~\cite{liu2010sentiment}. Sentiment analysis has a wide range of applications in several fields, such as social media analysis, product review analysis, public opinion monitoring, and user sentiment feedback analysis. Social media platforms for sentiment analysis can be used to track changes in user sentiment and public opinion to help companies and governments better understand public opinion. In addition, sentiment analysis can also be used to analyze product reviews to help manufacturers understand the strengths and weaknesses of their products and improve product design. Sentiment analysis is also widely used in opinion monitoring to help governments and organizations understand public reactions to major events promptly~\cite{feldman2013techniques}.

Sentiment analysis can be viewed as a categorization process that is divided into three main categorization levels: document-level, sentence-level, and aspect-level. Document-level SA aims to categorize an entire document into positive or negative sentiment, while sentence-level SA categorizes the sentiment of each sentence, first determining whether the sentence is subjective or not, and then determining the sentiment polarity. Aspect-level SA, on the other hand, categorizes sentiment for specific aspects of an entity, first identifying the entity and the relevant aspect. The above levels of sentiment analysis help in understanding sentiment expression and sentiment polarity in text~\cite{medhat2014sentiment}. Sentiment and subjectivity categorization is an important research area in the field of sentiment analysis and has been widely explored.

There are various methods for sentiment analysis, and the most common analysis methods are mainly based on rules and lexicons. The method relies on a predefined list of sentiment words as well as the frequency and contextual information of the sentiment words in the text~\cite{hutto2014vader}. For example, VADER (Valence Aware Dictionary and Sentiment Reasoner), which we use in this study, is a lexicon and rule-based sentiment analysis tool. The model was proposed by Hutto et al.~\cite{hutto2014vader} to specifically focus on the sentiments expressed in social media to address the unique challenges of social media content. The study first constructed a set of gold-standard lexical features specific to similar contexts of tweets and validated them through quantitative and qualitative methods. These lexical features are then combined with five generalized rules to capture grammatical and syntactic conventions of sentiment intensity. The results of the study show that in comparison with 11 other common sentiment analysis methods, the VADER model performs well in sentiment analysis, even outperforming individual manual raters, and also exhibits better generalization across different contexts, demonstrating the potential application of the VADER model in social media sentiment analysis~\cite{hutto2014vader}. Thapa utilized VADER to assess the sentiment of cybersecurity content on Twitter and Reddit, yielding an accuracy of 60\% for Twitter and 70\% for Reddit~\cite{thapa2022sentiment}. Given the textual similarities between Twitter, Reddit, and Discord, especially considering that Reddit and Discord both take on a forum-style social media format, it is reasonable to believe that VADER has significant potential for analyzing the emotional tone of the discussions on Discord in this study.

\subsection{Blockchain Governance}

Blockchain is a decentralized ledger system that employs synchronization mechanisms within a peer-to-peer database to enable transaction verification, recording, and distribution without intermediaries, thus eliminating the need to place trust in third parties~\cite{laatikainen_system-based_2023}. As blockchain technology matures, its applications have expanded into various domains, such as data sharing, item tracking, supply chain monitoring, power systems, and digital voting~\cite{mosley_towards_2022}. By enabling distributed trust, blockchain empowers a decentralized economy~\cite{zhang2022sok}.

In the realm of blockchain governance, the significance of effective decision-making mechanisms is paramount~\cite{laatikainen_system-based_2023,kiayias_sok:_2023,liu2024economics}. Robust governance systems play a pivotal role in the successful development and adaptability of blockchains, ensuring their resilience and ability to align with stakeholders' evolving needs and preferences~\cite{laatikainen_system-based_2023,kiayias_sok:_2023,mosley_towards_2022}. Effective governance mechanisms are instrumental in preventing untrustworthy behavior among participants and averting potential crises~\cite{laatikainen_system-based_2023}. Unlike centralized organizations led by a select few, blockchain platforms operate in a decentralized manner, striving to accommodate the diverse preferences of their participants~\cite{kiayias_sok:_2023}. However, the decentralized nature of blockchain often leads to governance challenges, such as issues related to participation levels, fairness in voting due to plutocratic systems, conflicts of interest, and transparency concerns~\cite{kiayias_sok:_2023,laatikainen_system-based_2023}. Addressing these governance issues effectively is imperative to harness the transformative potential of blockchain technology.

Previous studies have provided a comprehensive framework for understanding blockchain governance~\cite{laatikainen_system-based_2023,kiayias_sok:_2023,mosley_towards_2022}. Governance, in a broad context, entails the regulation of decision-making processes among participants to achieve common objectives that support the development, reinforcement, or continuity of social norms and institutions~\cite{laatikainen_system-based_2023}. Blockchain governance includes actions, systems, methods, and processes that facilitate decision-making, broadly categorized into technical and social means~\cite{laatikainen_system-based_2023}. Technical means predominantly support on-chain governance and automate decisions through various voting mechanisms, including smart contracts, DApp frameworks, and consensus algorithms~\cite{laatikainen_system-based_2023}. However, it is crucial to emphasize that technical means cannot solely bear the responsibility of governance decisions, as off-chain governance necessitates the involvement of social means~\cite{laatikainen_system-based_2023}. Social governance underscores formal and informal communication and collaboration among actors, with discussions held through forums and informal online voting on decisions serving as prominent examples~\cite{laatikainen_system-based_2023}. Furthermore, social media platforms serve as vital information conduits, influencing key stakeholders in blockchain governance~\cite{laatikainen_system-based_2023}.

While early studies explored various strategies to optimize blockchain governance, the growing significance of social factors, particularly public discourse, in influencing blockchain governance decisions has become increasingly evident. Filippi and Loveluck extensively examined blockchain governance structures, concluding that addressing social and political challenges in blockchain networks often requires more than technological solutions alone~\cite{defilippi:hal-01382007}. Kiayias and Lazos observed that many blockchains incorporate both on-chain and off-chain governance elements, utilizing forum discussions within their formal governance models~\cite{kiayias_sok:_2023}. Additionally, Rennie et al. conducted an ethnographic study of governance interactions within blockchain forums, shedding light on how activities in decentralized community spaces influence decision outcomes~\cite{rennie_toward_2022}. Therefore, there is a compelling need to comprehensively explore the social dimensions of blockchain governance. Specifically, it is imperative to investigate the social sentiment expressed through suggestions, debates, and complaints by participants and the broader public on voting platforms, as well as the open communication about system design through public media channels (e.g., blog posts, social media, podcasts). Such sentiment analysis can offer valuable insights to guide decision-makers in the realm of blockchain governance.

To facilitate this exploration, various sources of blockchain social discussion data have been identified in the existing literature, including Discord, Twitter, Reddit, Slack, and Github. Among these options, this paper employs discussion data from Discord, a widely used platform for discussing blockchain governance, as a dataset for analyzing the social aspects of blockchain governance.

\subsection{DAOs}
A decentralized autonomous organization (DAO) refers to a jointly-owned organization utilizing blockchain-based governance mechanisms that share a common goal. It operates independently without the need for a central authority or hierarchical management structure. Within a DAO, management and operational guidelines are encoded as smart contracts on the blockchain, and the organization achieves self-sufficiency, self-governance, and self-adaptation through distributed consensus protocols and a Token Economy Incentive system~\cite{8836488}. DAOs amalgamate the benefits of blockchain and decentralization, challenging conventional hierarchical management structures and substantially cutting down operational expenses for organizations~\cite{8836488}. In other words, DAO stands out as one of the most emblematic implementations of blockchain governance.

In theory, DAO members are motivated by profit sharing, encouraging each member to actively participate in DAO voting and governance, with the aim of enhancing community efficiency. However, in practice, this motivation often falls short. One reason is the inherent feature of decentralized autonomy, allowing DAO participants to exit at any time without the consent of others, especially when the governance is inefficient~\cite{laturnus2023economics}. Another factor is the tedium nature of the governance voting process, which drains the enthusiasm of community members, leaving them with limited energy to engage. For instance, the BitShares exchange has grappled with a problem of insufficient voter engagement, primarily stemming from the substantial workload associated with reviewing each proposal~\cite{neiheiser2020hrm}.

Adding to the voting dilemma is the alarming prevalence of pointless votes and delegations within many DAOs~\cite{daotimes2023centralization}. For instance, in \textit{Uniswap}, a staggering 88\% of voting power is concentrated in accounts holding fewer than ten tokens, and 47\% of the votes originate from accounts with just one token, despite the substantial requirements of 2.5 million and 40 million tokens, respectively, for proposal submission and approval~\cite{daotimes2023centralization}. This compounds the voting issue and further weakens the power of DAOs.

While the governance mechanisms of DAOs may appear decentralized in principle, the extent to which they achieve true decentralization in practice remains a matter of debate. In DAOs, the voting power of members is closely tied to the number of governance tokens they hold, with those holding more tokens wielding greater influence. This concentration of voting power means that influential token holders may prioritize their own interests at the expense of the broader community, potentially leaving others with little choice but to comply~\cite{laturnus2023economics}. Thus, for an ideal outcome, it is essential that each member of a DAO holds an equal number of tokens. However, numerous studies have shown a high degree of token centralization~\cite{laturnus2023economics}. Even in prominent DAOs like Compound, Fei, and \textit{Uniswap}, the majority of governance tokens are concentrated in the hands of a few, resulting in a limited degree of actual decentralization~\cite{daotimes2023centralization}. An illustrative case involves Build Finance DAO, where an individual consolidated a substantial number of governance tokens, thereby granting themselves absolute authority within the DAO. Subsequently, this individual proceeded to extract all the funds from the DAO~\cite{chohan2017decentralized}. Thus, there have been concerns about DAOs, with the suggestion that they might be nothing more than a facade crafted as a marketing tactic, concealing an authoritarian regime underneath.
So, how can we further enhance the governance mechanisms of DAOs? Research, as highlighted in Laturnus’s paper, has shown that active DAO communities tend to perform better~\cite{laturnus2023economics}. However, similar to the free rider problem we discussed earlier, the inherent characteristics of DAO mechanisms can incentivize community members to become passive participants.

To promote greater member engagement and improve the overall performance of DAO communities, it's essential to consider not only changes to the mechanisms themselves but also conduct a thorough social analysis. Given that members within DAOs have the capability to propose changes to the organization's rules~\cite{laturnus2023economics}, analyzing the sentiments expressed on social media and understanding how individuals perceive the governance of DAOs can play a pivotal role in fostering greater participation and optimizing the governance of DAOs.

\section*{Qualitative Analysis Result}
\label{qua}

We note that the results of sentiment analysis are too general, and the analysis results of different discussion communities may converge due to the limitations of sentiment analysis methods themselves. Machines seem unable to detect nuances in human language. Therefore, We conducted separate case studies of four Discord communities with a "governance" channel-\textit{Aave}, \textit{Uniswap}, \textit{Curve Dao}, and \textit{Yearn.finance}.

Based on the results of the quantitative sentiment analysis, we selected the day with the highest (most positive) and lowest (most negative) average sentiment scores for each channel, followed by one day with neutral sentiment, for a total of three days for manual review of the day's discussion text (see Table~\ref{tab:case}).
The results of the manual review exhibited some common problems. 

First, machines typically perform better at detecting neutral emotions, and in this case study, the machine's detection was accurate for all four randomly selected neutral emotions. Second, the machine's failure rate for detecting positive emotions is high. In this case, since there is a certain level of advertisement cooperation information in all forums, the advertisement publishers tend to use quite positive words to attract attention, and this part of the information is usually misrecognized as positive. Third, for the detection of negative emotions, there are two easily misdiagnosed aspects. On the one hand, the self-effacement of the text publisher, e.g., the use of \textit{"a dumb question"} in the negative example of \textit{Yearn.finance}, may lead to misrecognition by the machine. On the other hand, there are questions and doubts of users about objective problems, such as the negative examples of \textit{Uniswap} and \textit{Curve Dao}, which are only the users' explanations about objective problems and do not reflect subjective attitudes. In addition, VADER's method of determining the sentiment of an overall utterance by examining the polarity of specific words cannot determine the complex and subtle human emotions through the contextual context, which leads to the limitations of code in recognizing the sentiment of social media texts.

\begin{table*}[!htbp]
\footnotesize
\caption{Case study of DeFi protocols' discord discussion; Light pink color represents agree with machine's sentiment detection; Pink color represents disagree with machine's sentiment detection.}
\label{tab:case}
\centering
\begin{tabularx}{\textwidth}{
    |>{\centering\arraybackslash}p{1.25cm}
    |>{\centering\arraybackslash}p{1.4cm}
    |>{\centering\arraybackslash}p{1cm}
    |>{\centering\arraybackslash}p{1.75cm}
    |>{\centering\arraybackslash}p{7.8cm} 
    |>{\centering\arraybackslash}c|}
\hline
\rowcolor{lightgray} \textbf{DApp} & \textbf{Sentiment} & \textbf{Score} & \textbf{Date} & \textbf{Text} & \textbf{Agree/Disagree} \\
\hline
\rowcolor{pink} Aave & Positive & 0.98 & \texttt{2023.03.30} & Hello @everyone, I'm Ahmed from Canada, I have been looking at the granting review process and was wondering if I could DM someone on the granting committee to offer a tool that may help improve the granting process. I work with Ethelo - a collective decision-making platform and we have been working with some DAOs (e.g Gitcoin and BigGreen DAO) to help with collective decision-making for their granting process with our multi-criteria decision platform. Here is an example template, for those that are interested:https://granting.ethelo.net/. The platform is very flexible and can be configured to fit all sorts of decision matrices, and I think it might be useful to your granting process. Thank you! & Disagree \\
\hline
\rowcolor{lightpink} Aave & Negative & -0.91 & \texttt{2021.09.26} & I have to speak my mind because its upsetting.   PLEASE STOP SENDING PEOPLE TO TWITTER IN ORDER TO TAKE PART IN ANYTHING.. Thats a joke.. Why are these protocols demanding they use twitter in order to take part in any type of airdrop? Makes me upset why these people send us to the site that censors users, forces them to remove their own post (get in to the psychology behind that, reeducation)..Crypto is about leaving these censored platforms. Tryants who rule them! You have no idea how much I miss out on because i stopped using that site.. Its a damn monopoly & Agree \\
\hline
\rowcolor{lightpink} Aave & Neutral & 0 & \texttt{2023.05.24} & Hi, hop on the goverance forum of Aave. If it is about listing, you need to fill an AIP and ARC. https://docs.aave.com/governance/aips & Agree \\
\hline
\rowcolor{pink} Uniswap & Positive & 0.99 & \texttt{2023.06.16} & Dear  Uniswap, I hope this message finds you well. My name is Fred , and I represent a passionate community of blockchain enthusiasts who firmly believe in the transformative potential of decentralized finance (DeFi). I am reaching out to you today to explore a mutually beneficial collaboration that could revolutionize the DeFi landscape. As you may be aware, the recent years have witnessed a remarkable rise in the popularity of decentralized finance, with numerous innovative protocols offering a diverse range of financial services. However, despite this progress, we believe that there is a significant untapped opportunity to integrate Satoshi Chain into the existing DeFi ecosystem, taking it to new heights of efficiency, security, and scalability. Satoshi Chain, named in honor of the enigmatic creator of Bitcoin, seeks to build upon the foundations laid by Bitcoin and further enhance the capabilities of DeFi protocols. By incorporating Satoshi Chain into your existing infrastructure, your DeFi protocol can benefit from its unique features & Disagree \\ 
\hline
\rowcolor{lightpink} Uniswap & Neutral & 0 & \texttt{2023.07.28} & Hello there. Does anyone know why there's only the vote to deploy Linea visible on Agora, while the vote to deploy on Base is visible on Tally, Uniswap and Boardroom, but not on Agora? & Agree \\
\hline
\rowcolor{pink} Curve Dao & Positive & 0.95 & \texttt{2021.09.23} & Hi, Folks - I'm curious how Curve approaches marketing through governance proposals / community spend. From what I've seen on the governance pages, there is little to no focus on marketing. The reason I ask, I work with Blockworks and we're partnering with the Bankless Podcast folks to put on the largest DeFi conference to date. We're want to get together great DeFi communities, like Curve, to celebrate everything DeFi together. It's called Permissionless. What's the best way to kick off a discussion within the community? & Disagree \\
\hline
\rowcolor{pink} Curve Dao & Negative & -0.54 & \texttt{2021.03.26} & What do u mean by violating SRP here? & Disagree \\
\hline
\rowcolor{lightpink} Curve Dao & Neutral & 0 & \texttt{2021.05.31} & New Gauge Vote for ETH+, a diversified LSD basket. Curve: https://dao.curve.fi/vote/ownership/344. Forum: https://gov.curve.fi/t/proposal-to-add-eth-eth-to-the-gauge-controller/9272. Convex: https://vote.convexfinance.com/proposal/ & Agree \\
\hline
\rowcolor{pink} Yearn. finance & Positive & 0.96 & \texttt{2022.02.16} & @defiglenn (won't DM you)  and yearn team - I'm conducting some research for BUIDLweek at ETHDenver. The goal of the study is to begin to characterize the relationship between DAO governance models and contributor sentiment. I personally think yearn's governance is pretty cool (it got a great shout out today at the DAODenver talks) and would LOVE to have qualitative data from yearn's team.  I wrote a very simple 60 second survey that can be completed 100\% anon. Is it okay to put out a call for contributors to take part in the survey? Also, if anyone is on the ground in Denver, I'd love to meet up & Disagree \\
\hline
\rowcolor{pink} Yearn. finance & Negative & -0.51 & \texttt{2021.01.08} & Maybe a dumb question but where does proposal voting happen these days? Anyone have a link to the portal? & Disagree \\
\hline
\rowcolor{lightpink} Yearn. finance & Neutral & 0 & \texttt{2023.04.12} & yea, it was my understanding that only veYFI can vote now. ahh gotcha, majority of veYFI supply makes sense. I thought just whatever is majority, even if it was 1 veYFI, would pass a vote & Agree \\
\hline
\end{tabularx}
\end{table*}

\section*{Future Study}
\label{future}

\begin{figure}[!htbp]
    \centering
    \includegraphics[width=0.8\linewidth]{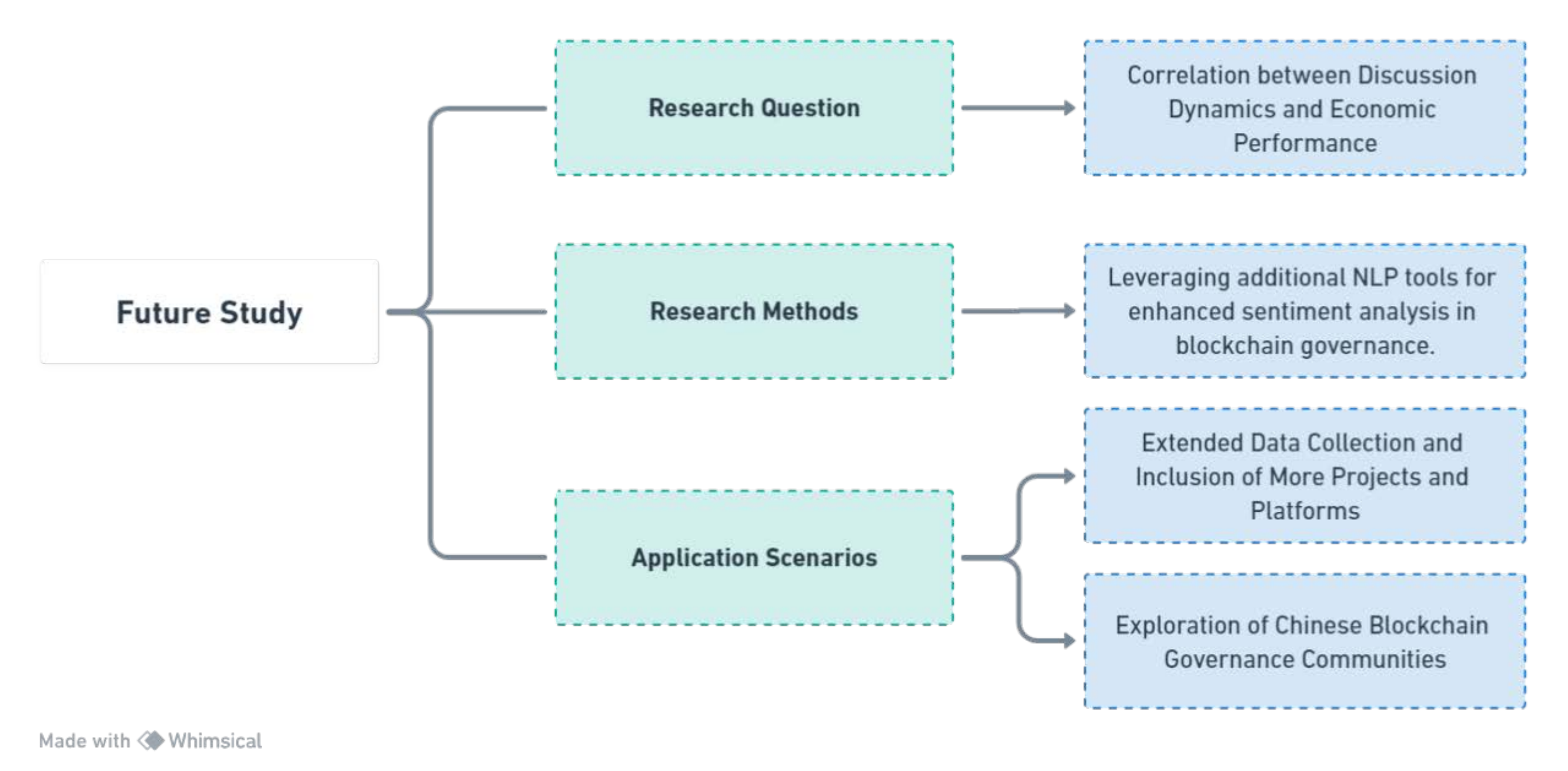}
    \caption{Future study logic flow: created by Whimsical.}
    \label{fig:future}
\end{figure}
As we conclude this investigation, we identify specific research questions, methodologies, and application scenarios that can deepen our understanding of blockchain governance and community dynamics. The logical flow of these future research directions is represented in Figure~\ref{fig:future}.

\subsubsection{Research Questions}
Building on blockchain DAO's economic performance indicators, such as the return on investment (ROI) of governance tokens~\cite{liu2022deciphering, liu2023cryptocurrency, zhang2022data, yu2023bitcoin}, transaction fees~\cite{liu2022empirical,zhang2023understand}, market exchange decentralization~\cite{zhang2022sok, ao2022decentralized,10.1145/3637305,chemaya2023uniswap,Zhang23,yan2024analyzing}, and cross-chain interoperability~\cite{interoperabilty}\footnote{Refer to the efforts by Dfinity in developing cross-chain interoperability tools: \url{https://github.com/dfinity/awesome-internet-computer}}, future studies could investigate whether a correlation exists between the intensity and sentiment of discussions within blockchain governance communities or DAOs and these economic and mechanic performance metrics. Understanding how community sentiment and engagement influence economic success could offer valuable insights to stakeholders and decision-makers.

\subsubsection{Research Methods} 

While VADER is popular due to its simplicity and efficiency, it is essential to recognize that lexicon and rule-based models may have limitations, particularly in handling complex variations in social sentiment analysis. In light of this, future research should consider exploring a broader array of sentiment analysis methods and NLP tools~\cite{fu2024dam,zhang2023machine}, including advanced AI-based models such as large language models (LLMs). We provide table~\ref{tab:nlp-tools} that lists nine possible NLP tools for future blockchain governance analysis and their related information. These sophisticated models may provide a more nuanced evaluation of the emotional dynamics within blockchain governance communities, offering enhanced accuracy and depth in the analysis of sentiment patterns.
\begin{table*}[!htbp]
\centering
\caption{NLP Tools for Future Blockchain Governance Analysis}
\label{tab:nlp-tools}
\small 
\begin{tabular}{p{1.7cm}p{1cm} p{3.5cm}p{3.5cm}p{1.5cm} p{1.6cm}}
\toprule
\textbf{Name} & \textbf{Docs} & \multicolumn{2}{c}{\textbf{Application Scenarios}} & \textbf{Language} & \textbf{Open Access} \\
\cmidrule(lr){3-4}
& & \textbf{Social Media} & \textbf{Blockchain} & & \\ 
\midrule
\midrule

TextBlob & \href{https://textblob.readthedocs.io/en/dev/}{URL} & Diyasa, I. Gede Susrama Mas, et al. (2021)~\cite{MasDiyasa_2021}& Hans, A., Mohadikar, K.R., Ekansh (2022) ~\cite{10.1007/978-981-16-8664-1_14}& Python & Yes: \href{https://github.com/sloria/textblob}{GitHub}\\

NLTK & \href{https://www.nltk.org/}{URL} & Elbagir, Shihab, Jing Yang (2019)~\cite{2020Sentiment}& Yang, Liuqing, et al. (2019)~\cite{10.1007/978-3-030-34083-4_13} & Python & Yes: \href{https://github.com/nltk/nltk}{GitHub} \\

Stanford CoreNLP & \href{https://stanfordnlp.github.io/CoreNLP/}{URL} & Kaur, Navjot, Arun Solanki (2018)~\cite{8442439}& Dulău, Tudor-Mircea, Mircea Dulău (2019) ~\cite{cryptocurrency}& Java & Yes: \href{https://github.com/stanfordnlp/CoreNLP}{GitHub} \\

spaCy & \href{https://spacy.io/usage/spacy-101}{URL} & Sharma, Mayank (2020)~\cite{9197829} & Coulter, Kelly Ann (2022)~\cite{coulter2022impact} & Python & Yes: \href{https://github.com/explosion/spaCy}{GitHub} \\

FastText & \href{https://fasttext.cc/docs/en/support.html}{URL} & Raihan, Muhammad Afif, Erwin Budi Setiawan (2022)~\cite{MuhammadAfifRaihan_ErwinBudiSetiawan_2022} & Kilimci, Zeynep Hilal (2020)~\cite{Kilimci_2020} & Python & Yes: \href{https://github.com/facebookresearch/fastText}{GitHub} \\

BERT & \href{https://arxiv.org/abs/1810.04805}{URL} & Heidari, Maryam, James H. Jones (2020) ~\cite{9298158}& Ider, Duygu, Stefan Lessmann (2022)~\cite{ider2023forecasting} & Python & Yes: \href{https://github.com/google-research/bert}{GitHub} \\

SentiWordNet & \href{https://github.com/aesuli/SentiWordNet/blob/master/papers/LREC10.pdf}{URL} & Miranda, Eka, et al. (2019) ~\cite{8843734}& Eisner, Lukas, Didier Sornette, Ke Wu (2019)~\cite{eisner2019interplay} & N.A. & Yes: \href{https://github.com/aesuli/SentiWordNet}{GitHub} \\

Flair & \href{https://flairnlp.github.io/docs/intro}{URL} & \href{uqua.duke.edu/programs/mqm-business-analytics/admissions-facts-dates}{Kula, Sebastian, et al. (2020)} & McMillan, Benjamin, et al. (2022)~\cite{AnalysisandComparison} & Python & Yes: \href{https://github.com/clips/pattern}{GitHub} \\

OpenAI-GPT & \href{https://platform.openai.com/docs/guides/gpt}{URL} & Wang, Zengzhi, et al.(2023) \cite{wang2023chatgpt} & N.A. & N.A. & No \\

\bottomrule

\end{tabular}
\end{table*}

\subsubsection{Application Scenarios}
\begin{table*}[!htbp]
\centering
\small
\caption{DeFi Protocols Information (Top 200 with Governance Coins and Discord - Data as of October 1, 2023)}
\label{tab:dapp}
\begin{tabular}{p{0.5cm}p{1.5cm} p{1.4cm}p{1.4cm}p{0.8cm}p{1.4cm}p{0.8cm}p{0.8cm}p{1.2cm}p{1.8cm}}
\toprule

Rank & DApp & Date of Genesis & Governance Token Symbol & Official Forum & Platforms & Social Media & White Paper & Etherscan & Economic Performance (Market Cap) \\

\midrule
\midrule

22 & Uniswap & 09/14/2020 & UNI & \href{https://gov.uniswap.org/}{URL} & \href{https://deepdao.io/organization/fe6aa70f-4877-4f6d-9c18-9a8cafc1fe28/organization_data/finance}{DeepDao}

\href{https://snapshot.org/#/uniswap}{Snapshot} & \href{https://twitter.com/uniswap/}{Twitter}

\href{https://www.reddit.com/r/UniSwap/}{Reddit}

\href{https://discord.com/channels/597638925346930701/}{Discord} & \href{https://uniswap.org/whitepaper-v3.pdf}{yes} & \href{https://etherscan.io/token/0x1f9840a85d5af5bf1d1762f925bdaddc4201f984}{URL}  & \$3,438,942,500 \\

35 & Maker & 11/25/2017 & MKR & \href{https://forum.makerdao.com/}{URL} & \href{https://deepdao.io/organization/c41f87df-35a6-4a37-82c4-62cd5a3a8c08/organization_data/finance}{DeepDao} & 
\href{https://classic.curve.fi/files/CurveDAO.pdf}{Twitter}

\href{https://www.reddit.com/r/MakerDAO/}{Reddit}

\href{https://discord.com/channels/893112320329396265/}{Discord} & \href{https://makerdao.com/en/whitepaper}{yes} & \href{https://etherscan.io/token/0x9f8f72aa9304c8b593d555f12ef6589cc3a579a2}{URL}  & \$1,396,627,300 \\

45 & Aave & 10/02/2020 & AAVE & \href{https://governance.aave.com/}{URL} & \href{https://deepdao.io/organization/9cc495be-6062-4b31-b4f8-424438ec1ed4/organization_data/finance}{DeepDao}

\href{https://snapshot.org/#/aave.eth}{Snapshot} & \href{https://www.reddit.com/r/Aave_Official/}{Twitter}

\href{https://www.reddit.com/r/Aave_Official/}{Reddit}

\href{https://discord.com/channels/602826299974877205/}{Discord} & \href{https://github.com/aave/aave-protocol/blob/master/docs/Aave_Protocol_Whitepaper_v1_0.pdf}{yes} & \href{https://etherscan.io/token/0x7fc66500c84a76ad7e9c93437bfc5ac33e2ddae9}{URL} & \$1,007,536,586 \\

80 & Curve Dao & 08/12/2020 & CRV & \href{https://gov.curve.fi/}{URL} & \href{https://deepdao.io/organization/802dad3e-2e3e-49d3-8fb5-4973c2e1751c/organization_data/finance}{DeepDao}

\href{https://snapshot.org/#/curve.eth}{Snapshot} & \href{https://twitter.com/curvefinance}{Twitter}

\href{https://discord.com/channels/729808684359876718/729819899685240893}{Discord} & \href{https://classic.curve.fi/files/CurveDAO.pdf}{yes} & \href{https://etherscan.io/token/0xD533a949740bb3306d119CC777fa900bA034cd52}{URL}  & \$455,976,605 \\

107 & Compound & 03/04/2020 & COMP & \href{https://www.comp.xyz/}{URL} & \href{https://deepdao.io/organization/52bf381b-79a8-4498-8504-41961beda494/organization_data/finance}{DeepDao}

\href{https://snapshot.org/#/comp-vote.eth}{Snapshot} & \href{https://twitter.com/compoundfinance}{Twitter}

\href{https://discord.com/channels/1135845501200240793/}{Discord} & \href{https://compound.finance/documents/Compound.Whitepaper.pdf}{yes} & \href{https://etherscan.io/token/0xc00e94cb662c3520282e6f5717214004a7f26888#tokenAnalytics}{URL} & \$325,514,927 \\

118 & PancakeSwap & 09/29/2020 & CAKE & \href{https://forum.pancakeswap.finance/}{URL} & \href{https://deepdao.io/organization/da9956dc-8a87-40c0-a066-b8991d67e574/organization_data/finance}{DeepDao}

\href{https://snapshot.org/#/cakevote.eth}{Snapshot} & \href{https://twitter.com/pancakeswap}{Twitter}

\href{https://discord.com/channels/897834609272840232/}{Discord} & \href{https://docs.pancakeswap.finance/}{yes} & \href{https://etherscan.io/token/0x152649eA73beAb28c5b49B26eb48f7EAD6d4c898}{URL} & \$265,883,084 \\

142 & Mask 

Network & 02/24/2021 & MASK & \href{https://we.mask.io/}{URL} & \href{https://deepdao.io/organization/fed90641-700b-477e-9179-72a19404bf58/organization_data/finance}{DeepDao}

\href{https://snapshot.org/#/masknetwork.eth}{Snapshot} & \href{https://twitter.com/realMaskNetwork}{Twitter}

\href{https://www.reddit.com/r/MaskNetwork/}{Reddit}

\href{https://discord.com/channels/757597809993056387/}{Discord} & \href{https://mask.io/faq?type=tutorials}{yes} & \href{https://etherscan.io/token/0x69af81e73a73b40adf4f3d4223cd9b1ece623074}{URL} & \$223,553,269 \\

151 & Ethereum Name Service & 11/09/2021 & ENS & \href{https://discuss.ens.domains/}{URL} & \href{https://deepdao.io/organization/228c8be1-9d1e-440c-8afb-a40d6ee480f2/organization_data/finance}{DeepDao}

\href{https://snapshot.org/#/ens.eth}{Snapshot} & \href{https://twitter.com/ensdomains}{Twitter}

\href{https://discord.com/channels/742384562646286509/}{Discord} & \href{https://docs.ens.domains/}{yes} & \href{https://etherscan.io/token/0xC18360217D8F7Ab5e7c516566761Ea12Ce7F9D72}{URL} &  \$214,443,516 \\

153 & Decred & 02/10/2016 & DCR & \href{https://forum.decred.org/forums/governance/}{URL} & Not 

Available & \href{https://twitter.com/decredproject}{Twitter}

\href{https://www.reddit.com/r/decred/}{Reddit}

\href{https://discord.com/channels/349993843728449537/}{Discord} & \href{https://www.allcryptowhitepapers.com/decred-whitepaper/}{yes} & Not 

Available & \$213,917,292 \\

155 & Aragon & 06/14/2017 & ANT & \href{https://forum.aragon.org/c/aragon-dao/83}{URL} & \href{https://deepdao.io/organization/18abd2d8-ff90-40cd-a889-4cc2fed78364/organization_data/finance}{DeepDao}

\href{https://twitter.com/AragonProject}{Snapshot} & \href{https://twitter.com/AragonProject}{Twitter}

\href{https://www.reddit.com/r/aragonproject/}{Reddit}

\href{https://discord.com/channels/672466989217873929/}{Discord} & \href{https://legacy-docs.aragon.org/aragon/readme}{yes} & \href{https://etherscan.io/token/0xa117000000f279d81a1d3cc75430faa017fa5a2e}{URL} & \$206,625,563 \\

168 & yearn.finance & 07/18/2020 & YFI & \href{https://gov.yearn.fi/}{URL} & \href{https://deepdao.io/organization/a1679a54-6e9a-4b03-b3da-fdcc7aeb28cd/organization_data/finance}{DeepDao}

\href{https://snapshot.org/#/ybaby.eth}{Snapshot} & \href{https://twitter.com/iearnfinance}{Twitter}

\href{https://www.reddit.com/r/yearn_finance/}{Reddit}

\href{https://discord.com/channels/734804446353031319/}{Discord} & \href{https://www.allcryptowhitepapers.com/wp-content/uploads/2020/12/YFI3.pdf}{yes} & \href{https://etherscan.io/token/0x0bc529c00C6401aEF6D220BE8C6Ea1667F6Ad93e}{URL} & \$179,651,224 \\

178 & Merit Circle & 11/03/2021 & MC & \href{https://gov.meritcircle.io/}{URL} & \href{https://deepdao.io/organization/00cdddf8-87e0-4e3f-9b36-f627133e0b24/organization_data/finance}{DeepDao}

\href{https://snapshot.org/#/meritcircle.eth}{Snapshot} & \href{https://twitter.com/MeritCircle_IO}{Twitter}

\href{https://discord.com/channels/851375133988749343/}{Discord} & \href{https://meritcircle.gitbook.io/merit-circle/}{yes} & \href{https://etherscan.io/token/0x949D48EcA67b17269629c7194F4b727d4Ef9E5d6}{URL} & \$157,189,222 \\

186 & Balancer & 06/23/2020 & BAL & \href{https://forum.balancer.fi/c/governance/7}{URL} & \href{https://deepdao.io/organization/11eb10d7-6493-4cfc-8255-cb0f2338fa67/organization_data/finance}{DeepDao}

\href{https://snapshot.org/#/balancer.eth}{Snapshot} & \href{https://twitter.com/Balancer}{Twitter}

\href{https://discord.com/channels/638460494168064021/}{Discord} & \href{https://balancer.fi/whitepaper.pdf}{yes}& \href{https://etherscan.io/token/0xba100000625a3754423978a60c9317c58a424e3D}{URL} & \$146,525,341 \\

\bottomrule
\end{tabular}
\end{table*}

\paragraph{Extended Data Collection and Inclusion of More Projects and Platforms:} 
Table~\ref{tab:dapp} encompasses information about the top 200 DeFi protocols with Discord forums and governance tokens, including market rankings, application names, genesis dates, governance token symbols, official forum, platforms, social media presence on Twitter, Reddit, and Discord, whitepaper availability, links from Ethereum to governance token smart contracts, and key economic performance indicators. To offer a more comprehensive perspective, future studies could extend the data collection period, include additional DeFi protocols and DAOs beyond the dApps on the Ethereum blockchain like the Internet Computer Protocols (ICP)~\cite{internetcomputer2023,liu2023economics}, and explore diverse communication platforms such as official governance channels, Reddit and Twitter.
\paragraph{Exploration of Chinese Blockchain Governance Communities:} 
Notably, several leading DeFi Discord communities including \textit{Uniswap}, \textit{Curve Dao}, Ethereum Name Service, and \textit{Yearn.finance} have established dedicated Chinese channels. Given China's significant standing in blockchain-related patent applications and the government's explicit support for blockchain industry development outlined in its "14th Five-Year Plan", delving into Chinese blockchain governance communities becomes imperative~\cite{cai_blockchain_2021}. Future research endeavors could systematically investigate Chinese Discord channels and other social media platforms like Weibo, the most popular Chinese microblogging social network, to unveil governance profiles and sentiment dynamics within the Chinese blockchain community. Employing Natural Language Processing (NLP) tools for Chinese text analysis can facilitate comprehensive semantic and sentiment analysis in this context. Notably, the use of SnowNLP\footnote{https://github.com/isnowfy/snownlp}, a well-established Python library for Chinese sentiment analysis, is recommended for its proven efficacy in deciphering the intricate nuances of Chinese semantic environments, especially for sentiment analysis on platforms like Weibo~\cite{li_study_2023,tang_evaluation_2020,zhou_sentiment_2022,el-hashash_analysis_2023,jia_emotional_2020}.

\balance

\end{document}